\documentclass[12pt]{article}
\usepackage[english]{babel}
\usepackage[latin1]{inputenc}

\usepackage[superscript,biblabel]{cite}
\usepackage{graphics} 
\usepackage{graphicx} 
\usepackage{float} 
\usepackage{subfig}
\usepackage{times}

\usepackage{etoolbox}
\newtoggle{withfigures}
\toggletrue{withfigures}
\usepackage{float} 
\usepackage{subfig}
\usepackage{array}
\usepackage{lscape}

\usepackage[letterpaper]{geometry}
\usepackage{vmargin}
\setmarginsrb{2cm}{2cm}{2cm}{2cm}%
{1cm}{0.5cm}{1cm}{1cm}   

\title{Synthesis and materialization of a reaction-diffusion French flag pattern\\
}

\author
{
Anton S. Zadorin$^{1,2}$, Yannick Rondelez$^{3,4}$, Guillaume Gines$^3$, Vadim Dilhas$^{1,2}$,  \\ 
Georg Urtel$^{5}$, Adrian Zambrano$^{1,2}$, Jean-Christophe Galas$^{1,2,\ast}$, Andr\'e Estevez-Torres$^{1,2,\ast}$\\
\\
\normalsize{$^{1}$Universit\'e Pierre et Marie Curie, Laboratoire Jean Perrin, 4 place Jussieu, 75005 Paris, France.}\\
\normalsize{$^{2}$CNRS, UMR 8237, 75005, Paris, France. $^{3}$LIMMS/CNRS-IIS, University of Tokyo,}\\
\normalsize{Komaba 4-6-2 Meguro-ku, Tokyo, Japan. $^{4}$Ecole sup\'erieure de physique et chimie industrielle,}\\
\normalsize{Laboratoire Gulliver, 10 rue Vauquelin, 75005, Paris, France. $^{5}$ Ludwig-Maximilians-Universität München,}\\ 
\normalsize{Fakultät für Physik, Amalienstraße 54, 80799 Munich, Germany}\\
\normalsize{$^\ast$To whom correspondence should be addressed;}\\
\normalsize{E-mail:   jean-christophe.galas@upmc.fr, andre.estevez-torres@upmc.fr.}
}

\date{}

\begin{document} 


\baselineskip24pt


\maketitle



\textbf{
During embryo development, patterns of protein concentration appear in response to morphogen gradients. These patterns provide spatial and chemical information that directs the fate of the underlying cells. Here, we emulate this process within non-living matter and demonstrate the autonomous  structuration of a synthetic material. Firstly, we use DNA-based reaction networks to synthesize a French flag, an archetypal pattern composed of three chemically-distinct zones with sharp borders whose synthetic analogue has remained elusive. A bistable network within a shallow concentration gradient creates an immobile, sharp and long-lasting concentration front through a reaction-diffusion mechanism. The combination of two bistable circuits generates a French flag pattern whose 'phenotype' can be reprogrammed by network mutation. Secondly, these concentration patterns control the macroscopic organization of DNA-decorated particles, inducing a French flag pattern of colloidal aggregation. This experimental framework could be used to test reaction-diffusion models and fabricate soft materials following an autonomous developmental program.
}

From a chemist's perspective, biological matter has the astonishing capability of self-constructing into shapes that are predetermined, robust to varying environmental conditions and remarkably precise in size and chemical composition at different scales. Living embryos, for instance, develop from a simple form into a complex one through a reproducible process called embryogenesis. The embryo is first structured chemically through pattern formation, a process during which out-of-equilibrium molecular programs generate highly ordered concentration patterns of $\mu$m to mm size\cite{Wolpert2011}. Subsequent developmental steps involve morphogenesis, where the embryo changes its shape, cell differentiation, in which cells become structurally and functionally different, and finally growth, resulting in an increase of the mass of the embryo\cite{Wolpert2011}. The emulation of such processes in non-living chemical systems adresses two important goals. First, in a reductionist perspective, it allows testing theoretical models describing the emergence of out-of-equilibrium material order in simplified experimental conditions\cite{Tompkins2014}. Second, from a synthetic standpoint, it enables the conception of a new way of making soft materials\cite{Yoshida1996, Inostroza-Brito2015} that one may call 'artificial development': materials that build themselves following a pre-encoded molecular program. 

The first developmental step, pattern formation, is currently interpreted in the light of two archetypal scenarios: Turing's instability \cite{Turing1952} and Wolpert's processing of positional information \cite{Wolpert1969}. The Turing scenario implies an initially homogeneous reaction-diffusion (RD) system that spontaneously breaks the symmetry, generating repetitive structures of wavelength $\sqrt{D\tau}$, where $D$ is a diffusion coefficient and $\tau$ a characteristic reaction time. In Wolpert's scenario, instead, the symmetry is already broken by a pre-existing morphogen gradient that is subsequently interpreted in a threshold-dependent manner, producing several chemically-distinct regions with sharp borders. Lewis Wolpert named this the French flag problem to illustrate the issue of creating three distinct regions of space ---the head, the thorax and the abdomen--- from an amorphous mass and a shallow concentration gradient (Fig. \ref{fig1}a). 
Historically, Turing's and Wolpert's scenarios have been considered mutually exclusive\cite{Green2015}, the first needing diffusion in contrast with the second. However, recent hypothesis suggest that both scenarios could be combined during development\cite{Green2015} and that diffusion could make the processing of positional information more robust\cite{Rulands2013,Quininao2015}. The experimental proof of the Turing mechanism in vivo has met constant criticism\cite{Green2015}, although recent work has provided new evidence\cite{Sheth2012, Economou2012}. In contrast, Wolpert's scenario is accepted in living embryos, possibly because it is more loosely defined. Well-known examples are provided by the gap gene system in \emph{Drosophila}\cite{Johnston1992} and by the sonic hedgehog-induced patterning of the vertebrate neural tube\cite{Briscoe2015}. Concerning non-living systems, Turing patterns were first demonstrated experimentally in 1990\cite{Castets1990} whereas synthetic systems capable of interpreting a morphogen pre-pattern have remained elusive.

Here, we draw inspiration from pattern formation during early \emph{Drosophila} development to address a synthetic challenge:  can one build a French flag pattern outside of a living organism? In a learning-by-doing approach\cite{Isalan2005, Loose2008, Padirac2013, Chirieleison2013, Semenov2014, Tayar2015} to this question our synthetic route combines reaction-diffusion with positional information, Turing and Wolpert ideas, suggesting that these two mechanisms are not antinomic. Furthermore, by considering pattern formation in the context of development as a blueprint for cell differentiation we ask a second question: may a self-organized chemical pattern influence the final structure of an initially homogeneous material? In doing this, we demonstrate that a soft material can be autonomously structured through an artificial developmental program.

\section*{Results}

The conversion of a shallow morphogen gradient into a concentration boundary that is both sharp and immobile requires a reaction network that interprets the gradient in a non-linear fashion. Diverse evidence \cite{Lewis1977, Francois2007, Lopes2008, Rulands2013} suggests that bistability is an essential property of such networks and that coupling bistability with diffusion provides immobile and robust fronts\cite{Lopes2008, Rulands2013, Quininao2015}. Our design for building a French flag pattern thus consists of two bistable networks that generate two RD fronts of concentration pinned in a gradient of a bifurcation parameter. 

\begin{figure}[h!]
\iftoggle{withfigures}
{
\begin{center}
\includegraphics[width=16 cm]{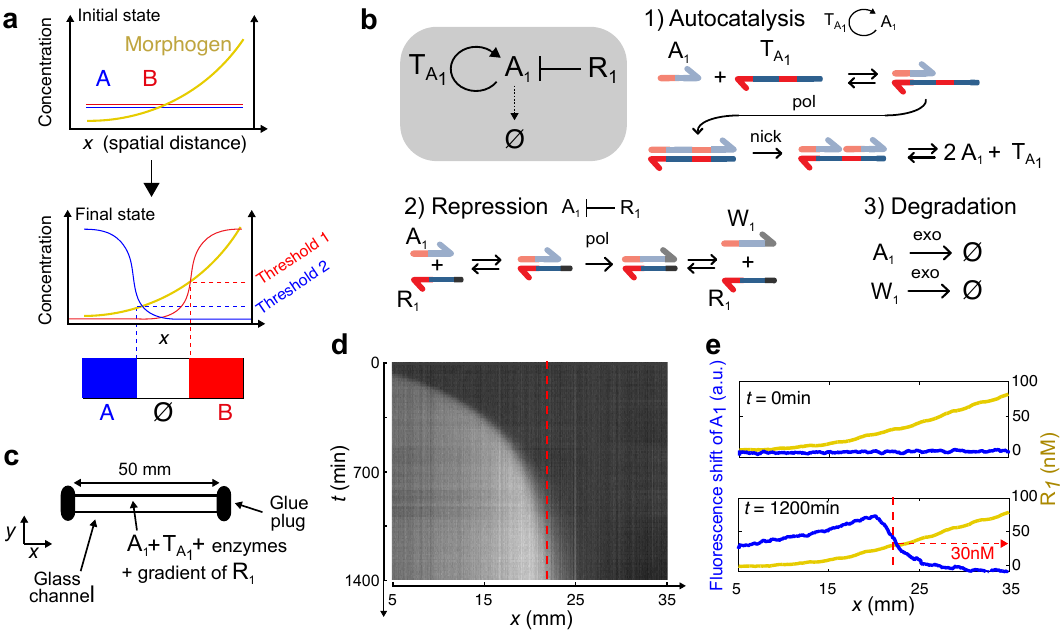}
\end{center}
}{}
{\small\caption{In a shallow gradient of morphogen, a bistable DNA network produces a Polish flag; a sharp and immobile concentration profile. \textbf{a}, Scheme of Wolpert's French flag problem, where a gradient of morphogen yields three chemically-distinct zones: blue, white and red. \textbf{b}, Molecular mechanism of a DNA-based bistable network where A$_1$ self-activation is supported by template T$_{A_1}$ and A$_1$ is repressed by R$_1$ and continuously degraded. Harpooned thick arrows are ssDNA where colors indicate sequence domains and light hue indicates complementarity. Straight black arrows denote chemical reactions. pol, nick and exo stand for polymerase, nicking enzyme and exonuclease, respectively. W$_1$ is a waste strand that cannot activate T$_{A_1}$. \textbf{c}, Sketch of the experimental setup. \textbf{d}, Kymograph of the fluorescence shift due to $A_1$ inside a capillary containing the network in \textbf{b} with homogeneous initial condition $A_1(x, t=0) = 1$~nM and pre-patterned with the gradient $R_1(x,t=0)$ as in \textbf{e}. The red dashed line indicates the stationary position of the profile. \textbf{e}, Profiles of $R_1$ (yellow) and the fluorescence shift due to $A_1$ (blue) along the channel at initial time and  after 1200 min. \label{fig1}}}
\end{figure}

DNA oligonucleotides are particularly well-suited to construct pattern-forming reaction networks.\cite{Chirieleison2013, Padirac2013, Scalise2014, Zadorin2015} The reactivity of the hybridization reaction obeys simple rules and a wide array of methods coming from biotechnology renders their synthesis, analysis and modification straightforward. We thus engineered a series of bistable networks using the PEN DNA toolbox, a molecular programming language designed to construct networks analogous to transcriptional ones but using only simple biochemical reactions.\cite{Montagne2011} This technology has recently been applied to construct out-of-equilibirum networks displaying oscillations\cite{Montagne2011, Fujii2013}, bistability\cite{Padirac2012} and traveling concentration waves.\cite{Padirac2013, Zadorin2015, Zambrano2015} Fig.~\ref{fig1}b depicts the simplest bistable network used here, with a first-order positive feedback loop and a non-linear repressor\cite{Montagne2016} (Supplementary Figs.~\ref{fig:MontBif1}-\ref{fig:MontBif3}).
The nodes of the network, A$_1$ and R$_1$, are respectively 11 and 15-mer single-stranded DNAs (ssDNAs). Self-activation is set by T$_{A_1}$, a 22-mer ssDNA template. Repression is encoded by promoting the degradation of A$_1$ with a threshold given by $R_1$\cite{Montagne2016} ---italized species names indicate concentration throughout the text. Three enzymes ---a polymerase, an exonuclease and a nicking enzyme---  catalyze the three basic reactions ---DNA polymerization, ssDNA degradation and the nicking of double stranded DNA (dsDNA) in the presence of the correct recognition sequence--- and dissipate free energy from a reservoir of deoxynucleoside triphosphates (dNTPs). In comparison with gene regulatory networks, template  T$_{A_1}$ plays the role of a gene, encoding information, A$_1$ and R$_1$ are analogous to transcription factors, as they promote or repress the activity of the template, and the three enzymes provide the metabolic functions homologous to the transcription-translation machinery.  Moreover, A$_1$ is continuously produced and degraded but the total template ---gene--- concentration is fixed. In contrast to networks in vivo, molecular interactions are well-known and the mechanism and kinetic rates can be precisely determined\cite{Montagne2011}. 

Our first goal was to create an immobile concentration front in the presence of a morphogen gradient, which we have called a Polish flag pattern. We performed patterning experiments within 5 cm-long sealed glass microchannels of $4\times0.2$~mm$^2$ cross-section (Fig.~\ref{fig1}c). A gradient of morphogen R$_1$  was generated along the longitudinal axis of the channel, noted $x$, by partially mixing two solutions with different $R_1$ by Taylor dispersion (Fig.~\ref{fig1}e and Supplementary Fig.~\ref{fig:gradFit}). The gradient was well approximated by an exponential decay $e^{(x-x_0)/l}$ with characteristic length $l = 2$~cm. Because diffusion is slow over long distances the gradient was stable over 50~h  (within 10\% at the center of the channel, as expected for the diffusion of a 17-mer ssDNA, see Supplementary Fig.~\ref{fig:gradFit}). Initially, the channel contained homogeneous concentrations of the three enzymes, dNTPs, $T_{A_1} = 25$~nM,  $A_1 = 1$~nM, and a gradient of $R_1$ in the range $0-200$~nM. Throughout this work, the concentration of the network nodes, here A$_1$, was related to fluorescence intensity using two methods (see Methods). Either by adding the DNA intercalator EvaGreen, for which the fluorescence signal is proportional to the concentration of dsDNA, or by labeling one template with a fluorophore that is quenched upon hybridization. In order to facilitate the interpretation of the data we represent in all figures the fluorescence shift, which is the absolute value of the difference between the fluorescence intensity at a given time and at initial time. As a consequence, at low concentration, the fluorescence shift is proportional to the concentration of the node species (see Methods). The fluorescence inside the channel was measured by recording time-lapse images with an inverted microscope. Fig.~\ref{fig1}d displays the spatio-temporal dynamics of the patterning process. A short, purely reactional initial phase generated a sharp profile of $A_1$ at a location corresponding to low morphogen concentration (Supplementary Fig.~\ref{fig:1Polish_flag_pT}). This profile later moved to the right through a RD mechanism, progressively decelerating until it stopped at the center of the channel at a position where $R_1^{RD} = 30\pm5$~nM. The front remained immobile up to 15 h and its characteristic width, defined as the decay length of a sigmoidal fit to the data, was $\lambda = 2$~mm, 10-fold sharper than the morphogen gradient (Fig.~\ref{fig1}d and Supplementary Fig.~\ref{fig:1Polish_flag_pT}). When, instead of the repressor, the autocatalyst template was used as the morphogen, the complementary Polish flag pattern was obtained (Supplementary Figs.~\ref{fig:1Bifurc_CB}-\ref{fig:1Polish_flag_CB}).

The gap gene network, which interprets the Bicoid morphogen gradient during the development of the \emph{Drosophila} blastoderm, is not composed of a unique self-activating and repressed node but of a series of them\cite{Manu2009} (Supplementary Fig.~\ref{fig:BicoidNetwork}). To demonstrate that our approach is capable of emulating the most basic type of such networks, we used one with two self-activating nodes, H and K, that repress each other\cite{Padirac2012} (Fig.~\ref{fig2}a and Supplementary Fig.~\ref{fig:PadiracBistable}) and recorded the patterning dynamics in a gradient of the Bicoid analogue, T$_{H}$ (Fig.~\ref{fig2}b-d, Movie S1), the template corresponding to autocatalyst H. At short times, a purely reactional phase created two independent and sharp fronts of H and K. Subsequently, during an RD phase, the fronts traveled in opposite directions until they collided in the middle of the channel. At this time, the two profiles partially overlapped and a slow phase made the two fronts go backwards, repelling each other, until reaching a steady-state where the overlap disappeared. 1-dimensional simulations with a 4-variable model (Supplementary Section~\ref{sec:simu} and Supplementary Fig.~\ref{fig:fitPadiracData}) displayed a similar behavior and suggested that this last phase was due to a slow synthesis of the repressors (Fig.~\ref{fig2}e-g). This patterning process was highly reproducible (Supplementary Figs.~\ref{fig:4nodesReproducibilityA}-\ref{fig:4nodesReproducibilityB}) and compatible with the immobilization of the morphogen gradient onto a surface (Supplementary Fig.~\ref{fig:Surface_1channel}).

To implement a French flag pattern with three chemically-distinct zones (Fig.~\ref{fig3}) we combined two orthogonal bistables, A$_2$ and A$_3$, (Supplementary Fig.~\ref{fig:bifFflag}) into a single network using two different approaches. We used a bifunctional morphogen bearing either both repressors, $\mathrm{R}_2-\mathrm{R}_3$, or one autocatalyst template and a repressor, $\mathrm{T}_{A_2}-\mathrm{R}_3$. In a gradient of $\mathrm{R}_2-\mathrm{R}_3$, a channel containing a uniform concentration of $\mathrm{T}_{A_2}$ and $\mathrm{T}_{A_3}$ generated a French flag pattern that  divided  space into three regions with different composition, $\textrm{A}_2 +\textrm{A}_3$, $\textrm{A}_2$ and $\emptyset$, for 100 min (Fig.~\ref{fig3}a, Supplementary Figs.~\ref{fig:FFlag2pT}-\ref{fig:FFlag2colors} and Supplementary Movie 2). By contrast, with a gradient of $\mathrm{T}_{A_2}-\mathrm{R}_3$ and  a uniform concentration of  $\mathrm{R}_2$  and $\mathrm{T}_{A_3}$ a different pattern separated the space into $\textrm{A}_3$, $\textrm{A}_2 +\textrm{A}_3$ and $\textrm{A}_2$ (Fig.~\ref{fig3}b).

In embryogenesis, pattern formation can induce tissue differentiation by providing localized chemical cues to pluripotent cells \cite{Wolpert2011}. This strategy could be used for the autonomous fabrication of artificial materials. As a proof of concept we coupled the patterns obtained above to the conditional aggregation of 1 $\mu$m diameter beads\cite{Mirkin1996} (Fig.~\ref{fig4}). Streptavidin-labeled beads were decorated with two types of biotin-labeled DNAs that had two different 12-mer ssDNA dangling ends, B$_i^l$ and B$_i^r$; for the pair of beads $i$, one has a \emph{left} and the other a \emph{right} strand. In the working buffer, the beads aggregated only in the presence of a linker strand L$_i$ complementary to both \emph{left} and \emph{right}  ssDNA portions (Supplementary Fig.~\ref{fig:BeadAggregationNoNetwork}). A capillary containing i) a homogeneous dispersion of beads B$_i^l$ and B$_i^r$, ii) a bistable network with node A$_j$ coupled to the linear production of L$_i$ and iii) a gradient of $\textrm{R}_j$, produced a Polish flag of bead aggregation (Fig.~\ref{fig4}c-e for $(i=1,j=1)$ and Supplementary Fig.~\ref{M161102} for $(i=2,j=3)$). 

It is straightforward to generate two pairs of beads, (B$_1^l$, B$_1^r$) and (B$_2^l$, B$_2^r$), each pair aggregating independently of the other in the presence of its own linker (L$_1$ or L$_2$, Supplementary Fig.~\ref{M161010_161102}-\ref{M161107_161109}). The French flag pattern in Fig.~\ref{fig3}a was thus materialized into three zones of space with distinct degree of aggregation (Fig.~\ref{fig4}f and Supplementary Fig.~\ref{M161108}): both pairs of beads aggregated / one pair of beads aggregated / no bead aggregated. In addition, bead aggregation brought two interesting properties. It allowed the visualization of RD patterns by eye (Supplementary Figs.~\ref{M161102} and \ref{fig:morphoflag_byEyes}) without the need of a fluorescence microscope and it froze the patterns into a state that was stable for at least one month, because aggregates precipitated to the bottom of the capillary (Supplementary Fig.~\ref{fig:morphoflag_byEyes}). In other words: a steady-state dissipative structure of DNA was converted into a kinetically-trapped stable structure of beads.

\section*{Discussion}

The object of synthesis in far from equilibrium chemistry is not anymore a molecular structure with desirable physico-chemical properties but a network of chemical reactions with particular dynamics. When these networks are coupled with some kind of transport ---diffusion is just one possibility\cite{Howard2011}--- a length scale naturally emerges, which allows to structure space chemically. The first implication of our work is to provide an experimental framework for the synthesis of far from equilibrium chemical structures. 
Indeed, among the limited amount of frameworks that have been proposed to synthesize RD patterns\cite{Vanag2009b, vanRoekel2015}, few of them are both biocompatible and programmable. By programmable we mean that one may rationally choose which reactants will react with what mechanism to create a desired pattern. The PEN DNA toolbox used here is naturally biocompatible and DNA sequence complementarity makes it programmable ---both for the topology of the network\cite{Montagne2011, Padirac2012, Padirac2013}, as shown in Fig. \ref{fig1}-\ref{fig4} and Supplementary Fig.~\ref{fig:networks}, and for the reaction and diffusion rates\cite{Zadorin2015}. Biocompatibility opens the 
way for the structuration of biomolecules and living cells with RD patterns. Programmability makes it suitable to synthesize new patterns, such as two-dimensional RD solitons\cite{Rotermund1991, Descalzi2013}, to probe experimentally the degree of complexity that may emerge from RD patterning. The versatility of this method allows conceiving a new way of processing materials inspired from embryo development: embed them with an autonomous developmental program and let them construct themselves. 

We may indeed  consider embryogenesis as an extremely precise and autonomous procedure to structure matter. In contrast with current fabrication methods, it is able to position chemicals with multiscale precision ---from tens of nm for cellular organelles to 10 $\mu$m for cells---, outstanding reproducibility and controlled dynamics without mechanical parts.  In this regard, RD mechanisms have already been used to process materials, notably for micro- and nanofabrication using precipitation reactions\cite{Grzybowski2005, Nakouzi2016}. Our approach significantly enlarges the complexity of the underlying reaction networks and, because information is encoded in DNA sequence, it could be coupled with directed evolution. 
If an autonomous fabrication method of this sort is once to find a real application it will need to build materials that are chemically diverse with a resolution at least comparable with the one in \emph{Drosophila} patterning, which is 10~$\mu$m. Recent work demonstrates that it is possible to couple DNA species with interesting chemistry such as DNA nanostructures\cite{Rothemund2006}, aptamers\cite{Franco2011}, hydrogels\cite{Lee2012}, chemical synthesis\cite{Gartner2001} or gene delivery\cite{Patwa2011}. Although the current $2$~mm resolution of our patterning chemistry is low, it is far from its theoretical limit. Indeed, RD patterning has been used for fabricating 300~nm objects\cite{Grzybowski2005}. The spatial resolution is determined by $\lambda\sim\sqrt{D/k}$, where $k$ is the degradation rate (Supplementary Fig.~\ref{fig:pTSimu}). Here $\lambda=2$~mm and taking $D = 1.8\times10^4~\mu$m$^2$/min for a 12-mer ssDNA\cite{Zadorin2015} at $45^\circ$C one finds $k \approx 5\times10^{-3}$~min$^{-1}$. This value is in good agreement with the measured degradation kinetics of species W$_2$ in the presence of R$_2$, with rates in the $3\times10^{-2}-3\times10^{-3}$~min$^{-1}$ range (Supplementary Fig.~\ref{fig:W1degradation}). Improving the resolution down to 10~$\mu$m thus requires decreasing $D$ and increasing $k$ each by a factor 100, which is not unreasonable (Supplementary Fig.~\ref{fig:W1degradation}). In addition, coupling RD with other sharpening mechanisms could increase resolution further:  the bead aggregation front was 4-fold sharper than the RD front, which  may be due to the cooperativity of bead aggregation\cite{Jin2003}.  This multi-level sharpening recalls hierarchical patterning mechanisms in vivo. Incidentally, it appears plausible to make 100 $\mu$m-scale gradients by either using  surface microprinting (Supplementary Fig.~\ref{fig:Surface_1channel}) or a self-generating diffusion-degradation mechanism\cite{Wartlick2009, Gines2017}.

The second implication of our work is that it may help understanding the physics of chemical patterning in living systems\cite{Giurumescu2009, Shvartsman2012}. Although our experimental system is an extreme simplification ---no cells, no molecular crowding, no forces---, it  integrates non-trivial microscopic interactions that are present in vivo ---intra and inter-molecular forces, DNA hybridization and protein-DNA recognition--- and all the time and spatial scales of reaction and diffusion are naturally included with actual molecules. It could thus provide an interesting  framework to perform experimental simulations instead of computer ones when these have limitations, for instance when noise becomes important.

\section*{Conclusion}

Our results demonstrate that DNA-based molecular programming is well-suited to engineer concentration patterns that emulate those observed during early embryo development. They indicate that the combination of a bistable reaction network with diffusion provides a simple engineer's solution to generate immobile concentration fronts that are both sharp and long lasting. Importantly, the simplicity of the method allowed us to record the patterning dynamics in real time, showing that a purely reactional initial phase is followed by a reaction-diffusion one. Our experimental model may help understanding the role of regulative and diffusive processes during development and suggests that relatively simple networks may have enabled patterning at an early stage of evolution.  Finally, by coupling programmable patterns with matter we have demonstrated a primitive autonomous chemical structuration of a material in one dimension. This approach could be exploited to fabricate soft materials following an autonomous developmental program.


\section*{Methods}

\begin{figure}[h!]
\iftoggle{withfigures}
{
\begin{center}
\includegraphics[width=16 cm]{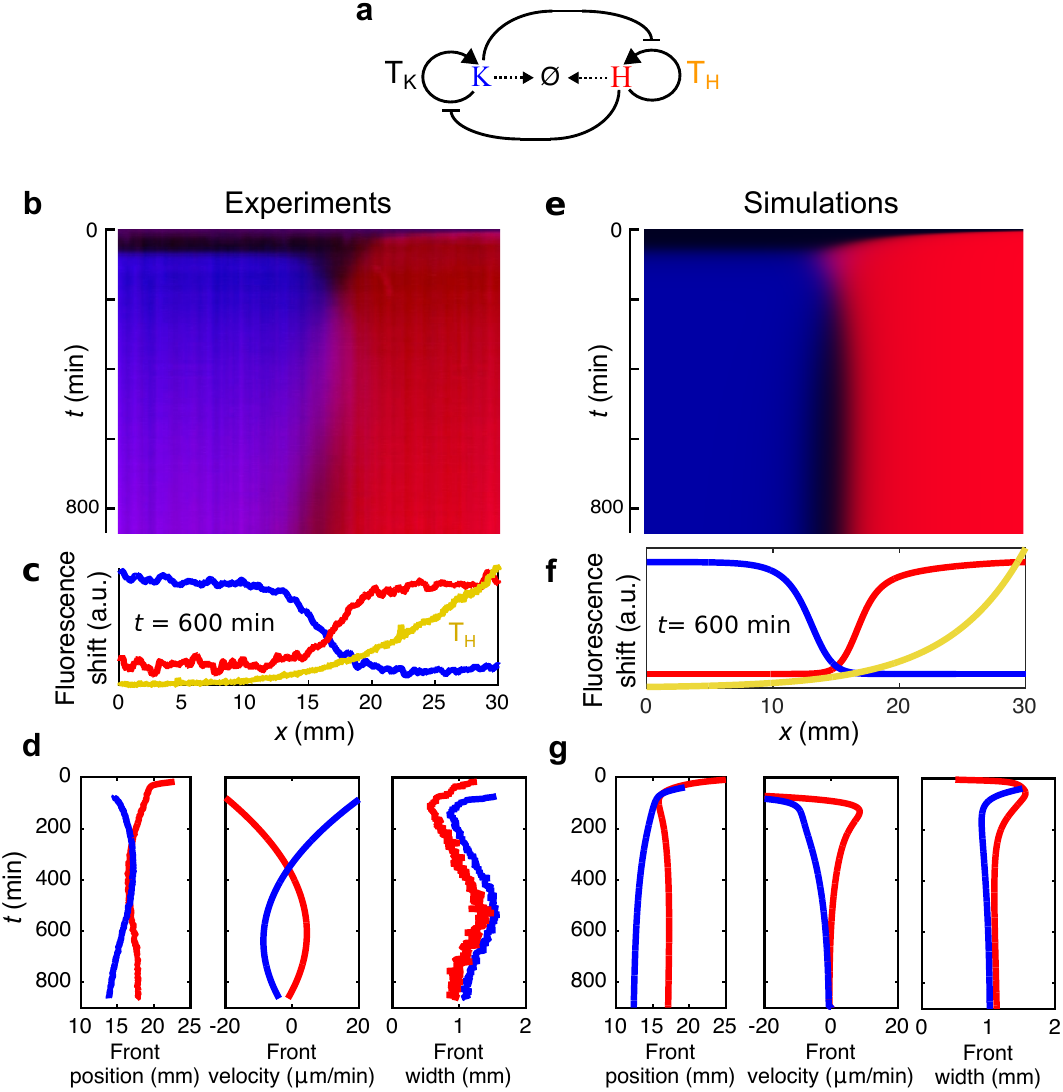}
\end{center}
}
{}
{\small\caption{A DNA-based network with two self-activating nodes that repress each other generates two immobile fronts that repel each other. \textbf{a}, Reaction network used here where H and K self-activate on their templates T$_H$ and T$_K$ and repress each other.  Experiments (\textbf{b}, \textbf{c}, \textbf{d}) and simulations (\textbf{e}, \textbf{f}, \textbf{g}) showing the kymograph of the fluorescence shift (\textbf{b}, \textbf{e}) for species H (red) and K (blue) inside a capillary containing the network in \textbf{a} and pre-patterned with a gradient of T$_{H}$ (yellow) and the fluorescence shift profiles at $t = 600$~min (\textbf{c}, \textbf{f}).  \textbf{d}, \textbf{g}, Front position, velocity and width as a function of time extracted from the kymograph (colors as in \textbf{c} and \textbf{f}). \label{fig2}}}
\end{figure}

\begin{figure}[h!]
\iftoggle{withfigures}
{
\begin{center}
\includegraphics[width=16 cm]{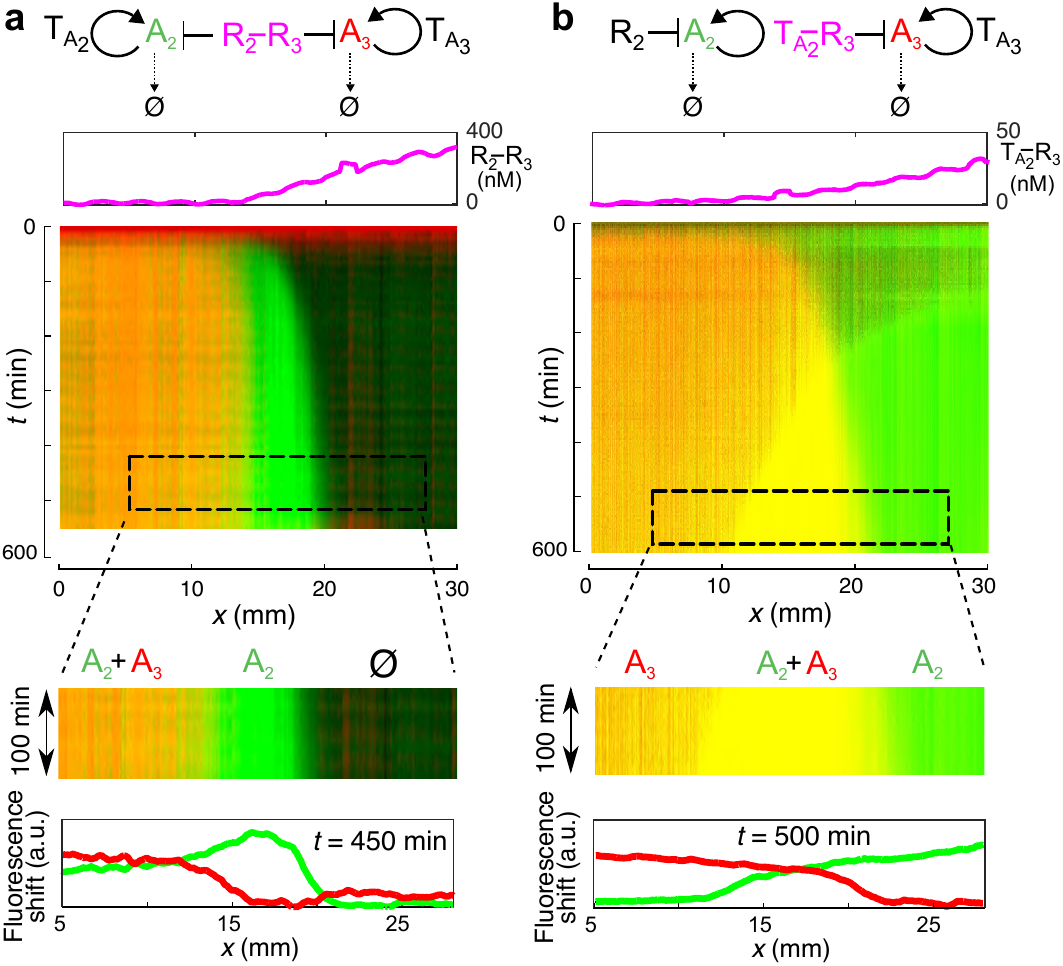}
\end{center}
}
{}
{\small\caption{The combination of two orthogonal bistable networks produces a French flag pattern of DNA concentration that can be simply reprogrammed. From top to bottom: network topology, initial morphogen gradient, kymograph and fluorescence shift profiles at steady state for two bistables coupled through either a double-repressor strand, $\mathrm{R}_2-\mathrm{R}_3$, \textbf{a}, or a template-repressor strand, $\mathrm{T}_{A_2}-\mathrm{R}_3$, \textbf{b}, each used as morphogen in the gradient. Dashed rectangles are zooms of the kymographs where the two French flag patterns were stationary.\label{fig3}}}
\end{figure}

\begin{figure}[h!]
\iftoggle{withfigures}
{
\begin{center}
\includegraphics[width=13cm]{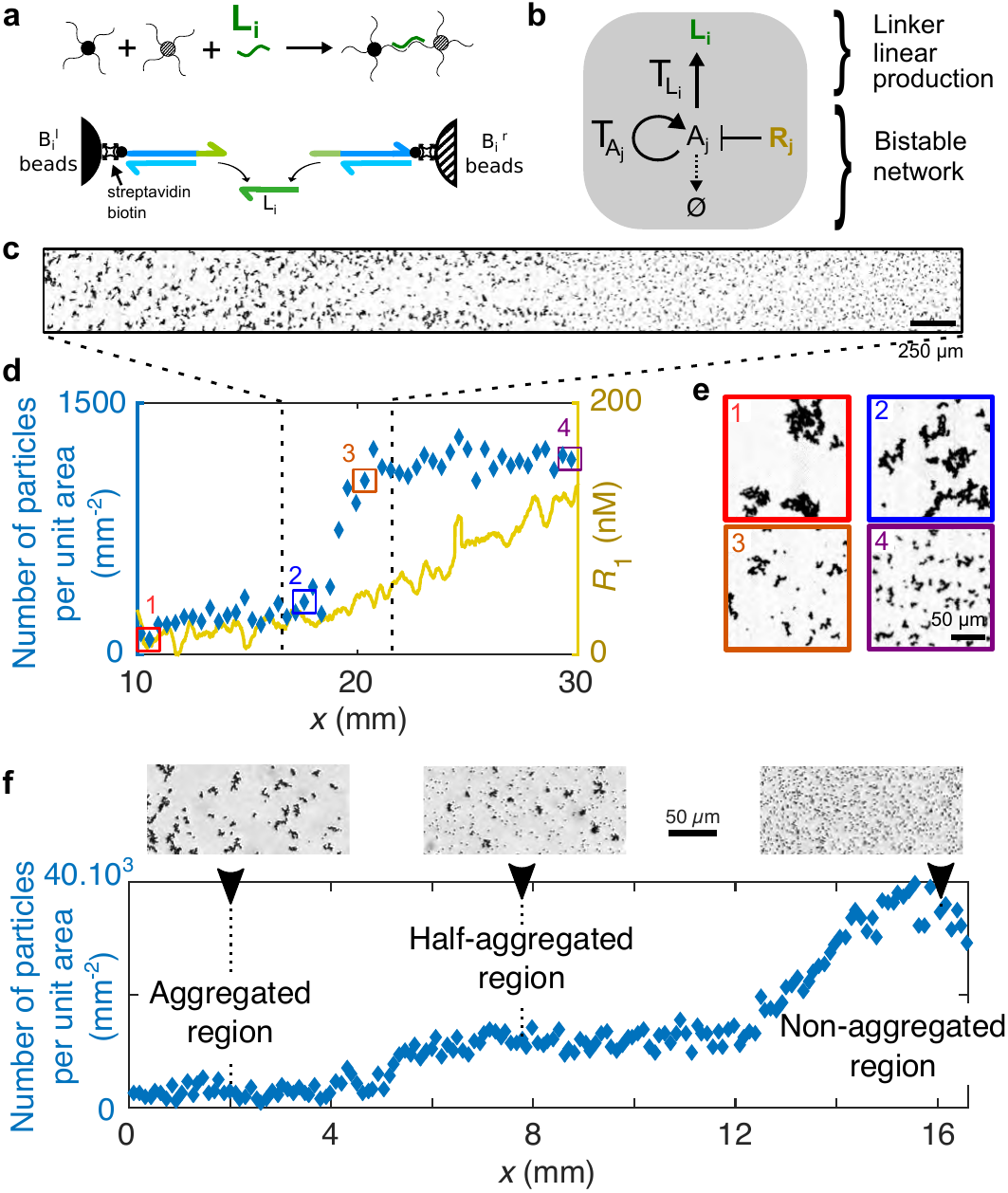}
\end{center}
}
{}
{\small\caption{Materialization of a Polish and a French flag pattern with conditional bead aggregation. \textbf{a}, Sketch of the mechanism of bead aggregation where a pair $i$ of 1~$\mu$m beads decorated with two different DNA constructs (black and gray disks) are aggregated in the presence of linker strand L$_i$. \textbf{b}, Scheme of the reaction network motif used to couple a bistable network based on species A$_j$ with the linear production of L$_i$ supported by template T$_{\mathrm{L}_i}$. \textbf{c-e}, Polish flag pattern of bead aggregation obtained after 40 h in a channel containing the beads $i=1$ and a network $j=1$ with an initial gradient of $R_1$. \textbf{c}, Brightfield image at the center of the channel. \textbf{d},  Number of particles per unit area (blue diamonds, left axis) and initial concentration of $R_1$ (yellow line, right axis) along the longitudinal axis of the channel. The colored squares indicate the positions at which the brighfield images in \textbf{e}  were recorded. The dashed lines correspond to the position where \textbf{c} was recorded. \textbf{f}, French flag pattern of bead aggregation obtained after 40 h in a channel containing the beads $i=(1, 2)$ and two networks $j=(1,3)$ with initial gradients of $R_1$ and $R_3$. \label{fig4} }
}
\end{figure}

DNA strands were purchased from Biomers (Ulm, Germany). The Bst DNA polymerase large fragment and the two nicking enzymes (Nb.BsmI and Nt.BstNBI) were purchased from New England Biolabs. The recombinant \emph{Thermus thermophilus} RecJ exonuclease was produced as described\cite{Wakamatsu2010}. Experiments were performed at 45 °C for the Polish and French flag generating networks and at 42 °C for those in Fig.~\ref{fig2}. Details on the the sample preparation, the DNA sequences, the experimental conditions, the data analysis and the simulations are provided in the Supplementary Information.

\paragraph{Bead suspension.} 1~$\mu$m diameter, streptavidin-coated, paramagnetic beads (Dynabeads MyOne C1, Invitrogen) were functionalized with two types of biotinylated DNA constructs, as described\cite{Leunissen2009}, making a pair of beads, (B$_i^l$, B$_i^r$). Each construct consisted of a 49 bp-long dsDNA backbone  terminated with a 12 bases-long single stranded sticky end. The construct corresponding to B$_i^l$ (resp. B$_i^r$) was biotin-labeled on the 5' end (resp. 3' end) and the corresponding sticky end was on the 3' side (resp. 5' side). 

\paragraph{Measurement of DNA concentrations.} DNA concentrations were measured by fluorescence. To measure the concentration of autocatalytic nodes A$_i$, H and K we used two different strategies. i) All experiments involving A$_i$ contained EvaGreen (Biotium), an intercalating dye which fluorescence is proportional to the concentration of dsDNA. ii) Some template strands were labeled with fluorescent dyes on their 3' or 5' ends (see Supplementary Table~\ref{tableSeqs} for details) which fluorescence was quenched specifically by the corresponding complementary strands. For experiments with networks involving a single node (Fig.~\ref{fig1}) we used EvaGreen. For networks involving two nodes we either used two orthogonal N-quenching dyes (Fig.~\ref{fig2}) or EvaGreen and an orthogonal N-quenching dye (Fig.~\ref{fig3}). The concentration of morphogen was measured either by using a fluorescently-labeled strand (Fig.~\ref{fig2}) or by adding a cascade blue-dextran M$_w = 3000$~Da  (Thermo Fisher Scientific)  (Fig. \ref{fig1} and \ref{fig3}-\ref{fig4}), and recording its fluorescence in real time. This fluorescent dextran has a diffusion coefficient similar to the DNA templates (Supplementary Fig.~\ref{fig:gradFit}).  All kymographs and plots display the fluorescence shift, i.e the absolute value of the difference between the fluorescence intensity at a given time and at initial time. The fluorescence shift is proportional to the concentration of the species of interest  if this remains below the concentration of its complementary strand.

\paragraph{Generation of the morphogen gradient.} Spatiotemporal experiments were performed within 50~mm $\times$ 4~mm $\times$ 0.2~mm glass capillaries (Vitrocom USA). One half of the capillary was loaded with the reaction solution without the morphogen strand and the other half with the same solution supplemented with the morphogen at 200 nM.  To create the gradient the two solutions were mixed by applying a hydrodynamic flow along the capillary axis using a micropipette and a custom-made PDMS connector. 15 up-and-down pumps of 12.5 $\mu$L yielded a shallow morphogen gradient along the entire capillary length spanning between 0 and $\sim$100 nM. Subsequently, the capillaries were sealed with 5-minutes Araldite epoxy and glued over a $5\times7.5$~cm glass slide.

\paragraph{Microscopy.} The fluorescence along the capillary was recorded on a Zeiss Axio Observer Z1 automated epifluorescence microscope equipped with a Tokai Hit thermo plate, an Andor iXon Ultra 897 EMCCD camera and a $2.5\times$ objective and  controlled with MicroManager 1.4. For each capillary, 16 contiguous 3.17$\times$3.17 mm$^2$ images were recorded automatically every 1 to 10 minutes.  Multi-color fluorescence microscopy was used to record the concentration of different DNA species over time. Images of the beads were acquired in bright field with a 10 or a $40\times$ objective.

 \paragraph{Data treatment.} The raw data were treated with ImageJ / Fiji (NIH) and Matlab (The Mathworks). The 16 images making one capillary were stitched together and corrected from inhomogeneous illumination. To obtain the kymographs, the images were averaged along the width of the capillary ($y$ axis) and the corresponding profiles stacked over time. These profiles were further smoothed by performing a moving average along $x$, normalized and fitted to a sigmoid function $f(x)=\frac{1}{1+e^{(x-x_0)/\lambda}}$. The front position corresponds to $x_0$ and its width to $\lambda$.  To determine the front velocity, the data of front position over time were fitted by a polynomial function to reduce noise. The velocity was calculated as the time derivative of this polynomial fit. 
To plot the bead aggregation profile along the capillary, the corresponding brightfield images were first binarized and the number of particles counted in areas 100 $\mu$m wide. A particle here indicates either an aggregate or a bead: the number of aggregates was thus inversely proportional to the number of detected particles.

 \paragraph{Data availability.} All data generated or analysed during this study are included in this published article (and its supplementary information files). The raw  datasets and the plasmid for expressing ttRecJ are available from the corresponding authors on reasonable request.

\section*{Acknowledgments}

We thank E. Frey for insightful discussions, A. Vlandas for help with gradient generation and B. Caller and D. Woods for comments on the text. Supported by European commission FET-STREP (Ribonets), by ANR jeunes chercheurs program (Dynano), by C'nano Ile-de-France (DNA2PROT) and by Ville de Paris Emergences program (Morphoart). Correspondence and requests should be addressed to A.E.-T. (andre.estevez-torres@upmc.fr) or J.-C.G (jean-christophe.galas@upmc.fr).

%

\section*{Author contributions}

A.S.Z., J.-C.G. and A.E.-T. performed most experiments and analyzed the data. Y.R. and G.G. designed the network in Fig.~\ref{fig1} and J.-C.G. and A.E.-T. designed the networks in Figs.~\ref{fig3}-\ref{fig4}. A.Z. and V.D. set up the bead experiments. G.U. performed critical control experiments. All the authors discussed the results. J.-C.G., A.S.Z., Y.R. and A.E.-T. designed research and J.-C.G. and A.E.-T.  wrote the manuscript.

\section*{Competing financial interests}

The authors declare no competing financial interests.

\section*{Table of contents summary}

During embryogenesis patterns of protein concentration appear in response to morphogen gradients, providing spatial and chemical information that directs the fate of the underlying cells. Here, this process is emulated with DNA-based non-living matter and the autonomous  structuration of a synthetic material is demonstrated.

 \newpage

\section*{Figure captions}

\setcounter{figure}{0}


\newpage

\begin{center}
{\LARGE Supplementary materials}
\end{center}
\newpage
\section{Materials and methods}
        
        \subsection{Preparation of solutions}
        
       Two types of bistable networks were used throughout the Main Text. A single species production with degradation network (Figures 1, 3 and 4) and a two species mutual inhibition network (Figure 2). They will be called, respectively, 1-species and 2-species bistable networks. The 1-species and 2-species bistable networks depend on two different nicking enzymes and thus the associated buffers slightly differ.
       
       \subsubsection{Components common to both buffers}
        
        1$\times$ thermopol buffer (NEB, containing 20~mM Tris-HCl, 10~mM (NH$_4$)$_2$SO$_4$, 10~mM KCl, 2~mM MgSO4, 0.1\% Triton X-100, pH 8.8@25°C),
        1~g/L synperonic F108 (Sigma Aldrich), 
        3~mM Dithiothreitol (Sigma Aldrich), 
        50~mg/L Bovine Serum Albumin (NEB), 
        2$\mu$M Netropsin (Sigma Aldrich), 
        50~$\mu$M NaCl.
        
        \subsubsection{1-species bistable buffer}
        
         In addition to the components common to both buffers, the 1-species bistable buffer contains: 
          \begin{itemize}
          \item MgSO$_4$ 6~mM,
         \item deoxynucleotidetriphosphates  (0.8~mM of each dNTP) (NEB),
          \item Bst DNA polymerase large fragment (pol) (NEB) 0.1\% of the 8,000 units/mL stock solution, 
         \item \emph{Thermus thermophilus} RecJ exonuclease (\emph{tt}RecJ) expressed in house as described in \cite{Wakamatsu2010}, 12.5~nM (1\% of the stock solution),
         \item Nb.BsmI nicking enzyme (nick) (NEB), 5\% of the 10,000 units/mL stock solution.
         \item If templates bearing a biotin in the 5' end were used, the buffer was supplemented with streptavidin at 200 nM in binding sites.
         \end{itemize}
         
         All experiments involving the 1-species bistable network were performed at 45°C.

\subsubsection{2-species bistable buffer}
        
          In addition to the components common to both buffers, the 2-species bistable buffer contains:        
          \begin{itemize}
          \item Tris-HCl 25~mM, pH 8,
          \item MgSO$_4$ 5~mM
          \item dNTPs (0.2~mM each) (NEB),      
          \item Bst DNA polymerase large fragment (pol) (NEB) 0.2\% of the 8,000 units/mL stock solution,
          \item \emph{tt}RecJ 16~nM (1.33\% of the stock solution),
          \item Nt.BstNBI nicking enzyme (nick) (NEB), 1\% of the 10,000 units/mL stock solution.
           \end{itemize}
                 
          All experiments involving the 2-species bistable network were performed at 42°C.
      
        Note that the activity of both nicking enzymes occasionally changed from batch to batch, thus their concentrations were adjusted according to independent assays.

\subsubsection{Particle suspension}

\begin{figure}[H]
 	\centering  
 	\includegraphics[width=\linewidth]{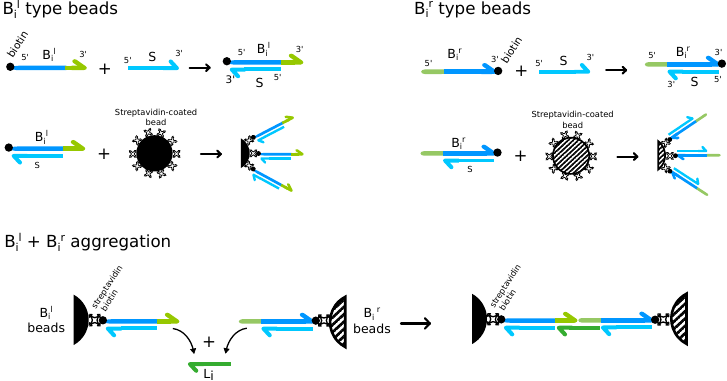}
 	\renewcommand{\figurename}{Supplementary Figure}\caption{Scheme of the DNA-functionalized beads. 1~$\mu$m in diameter, streptavidin coated beads are linked to biotinylated DNA constructs. Both constructs for B$_i^l$ and B$_i^r$ bead types consist of a rigid, double-stranded backbone, terminated by a single-stranded sticky end. B$_i^l$ and B$_i^r$ are designed to aggregate in the presence of the linker strand L$_i$.\label{fig:beadScheme}}
 \end{figure}

          The DNA-functionalized colloids used in this work are a slight variation of the design described in \cite{Leunissen2009}. We used 1~$\mu$m diameter, streptavidin-coated  (5~$\mu$M biotin binding sites and 10 mg/mL of beads) paramagnetic beads (Dynabeads MyOne C1, Invitrogen). The beads were decorated with two different biotinylated DNA constructs  making two different beads, called B$_i^l$ and B$_ir$ (Supplementary Figure \ref{fig:beadScheme}). Each construct consists of a 49 bp-long dsDNA backbone (S-S*) terminated with a 12 bases-long single stranded sticky end. The construct corresponding to B$_i^l$ (resp. B$_i^r$) was biotin-labeled on the 5' end (resp. 3' end) and the corresponding sticky end was on the 3' side (resp. 5' side). Such a design implies that the DNA-decorated beads will be stable in solution when mixed, unless the linker strand complementary to the sticky ends of each bead type is present.  The subscript $i$ indicates that the pair of beads aggregate with linker strand L$_i$. L$_i$ has 2 extra A bases on its 3' end to impede polymerase extension and the subsequent strand displacement of S on B$_i^l$ beads. The grafting protocol of DNA to get the two types of beads presented in Supplementary Figure \ref{fig:beadScheme} requires the following steps:

\begin{enumerate}
\item Two solutions with the two DNA constructs  (B$_i^l$ + S) and (B$_i^r$ + S) are prepared at 8~$\mu$M final concentration of strands in the suspension buffer (10~mM phosphate, 50~mM NaCl and 0.1\% w/w Pluronic surfactant F127, Sigma-Aldrich) and kept for at least 15 min at room temperature before use.            
\item  During this time, the beads are rinsed 3 times in the suspension buffer and split into two aliquots. 

 \item The bead supernatant is removed and the corresponding DNA construct solution is added to each bead aliquot and incubated at room temperature for 30 min under gentle mixing. The final concentration of the beads is 10 mg/mL. 
\item The two beads solutions are mixed and rinsed 3 times in the suspension buffer, then maintained at 55°C for 30 min and rinsed again 3 times, to eliminate the non-grafted strands. The bead stock concentration is kept at 10 mg/mL. The solution is stored at 4°C. 
\end{enumerate}

          Note that before each experiment, the bead stock solution is heated again at 55°C for 30 min. The bead solution is used at a concentration ranging from 0.1 to 0.5 mg/mL in the final reaction mix.
          
 \subsubsection{DNA concentrations for the experiments presented in the Main Text}

           \begin{table}[H]
           	\renewcommand{\tablename}{Supplementary Table}\caption{\label{table2} Experimental conditions used for the experiments presented in the Main Text Figures.}
           	\begin{center}
           		\begin{tabular}{|c | m{5cm} | m{8cm} | }
           			\hline
           			 Fig. & Network topology & Initial  conditions\\
           			\hline
           			1	& \includegraphics[width=3cm]{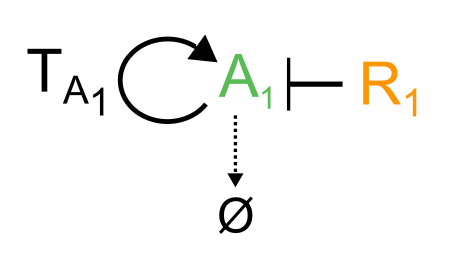} &
           			 T$_{A_1}=25$~nM, $A_1=1$~nM, $R_1=0-400$~nM\\	 
            		\hline
           		    2	& \includegraphics[width=4cm]{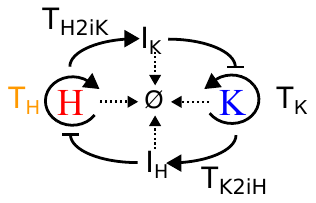} & $T_{H}=0-200$~nM, $T_{K}=20$~nM, $T_{H2R_K}=20$~nM, $T_{K2R_H}=20$~nM, $I_K=0$~nM, $I_H=0$~nM, $H=0.5$~nM, $K=0.5$~nM\\
           		    \hline           		    
           		    3A	& \includegraphics[width=5cm]{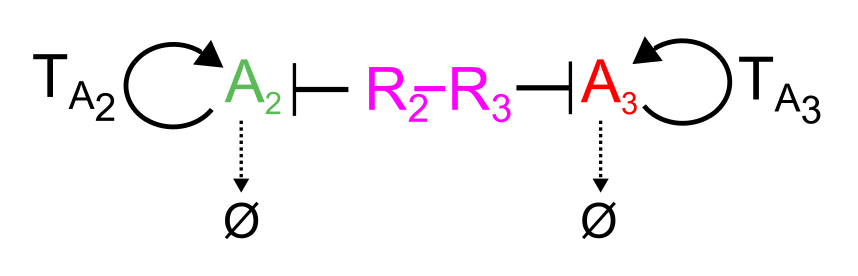} &  $T_{A_2}=75$~nM, $T_{A_3}=25$~nM, $A_1=10$~nM\textsuperscript{$\dagger$}, $A_3=10$~nM, $R_2-R_3=0-400$~nM, $T_{A_2}=25$~nM\\
           		    \hline
           		    3B & \includegraphics[width=5cm]{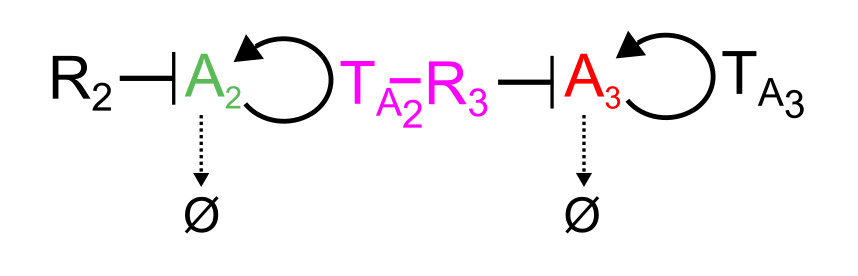} & $R_2=50$~nM, $T_{A_3}=25$~nM, $A_1=10$~nM\textsuperscript{$\dagger$}, A$_3$=10~nM, $T_{A_2}-R_3=0-75$~nM\\          		    
           		    \hline
           		    4C-E	& \includegraphics[width=3cm]{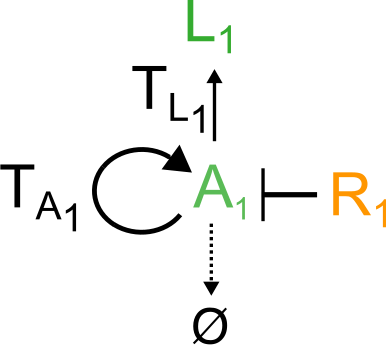} &  $T_{A_1}=25$~nM, $T_{L}=50$~nM,  $A_1=10$~nM, $R_1=0-400$~nM, $L_1=0$~nM\\
		    \hline
		    4F	& \includegraphics[width=5cm]{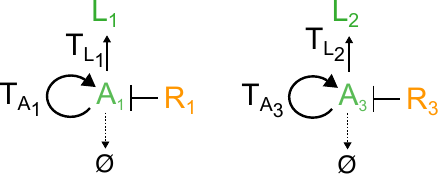} &  $R_1=0-400$~nM, $R_3=0-150$~nM, $A_1=A_3=10$~nM, $T_{A_1}=T_{A_3}=25$~nM, $T_{L1}=T_{L2}=50$~nM\\             		    
           			\hline
           		\end{tabular}
           	\end{center}
           \end{table}
 \textsuperscript{$\dagger$}  The node of the network corresponds well to species A$_2$. However it was initiated by species A$_1$ because the input side of templates T$_{A_1}$ and  T$_{A_2}$ are closely related and both can be triggered by species  A$_1$.

        \subsection{DNA sequences and topology of the networks}
        The DNA oligonucleotides were purchased from Biomers with HPLC purification.
        All the reaction networks are presented in Supplementary Figure \ref{fig:networks} and the sequences listed in Supplementary Table~\ref{tableSeqs}.

        \begin{figure}[H]
        	\centering  
        	\includegraphics[width=14cm]{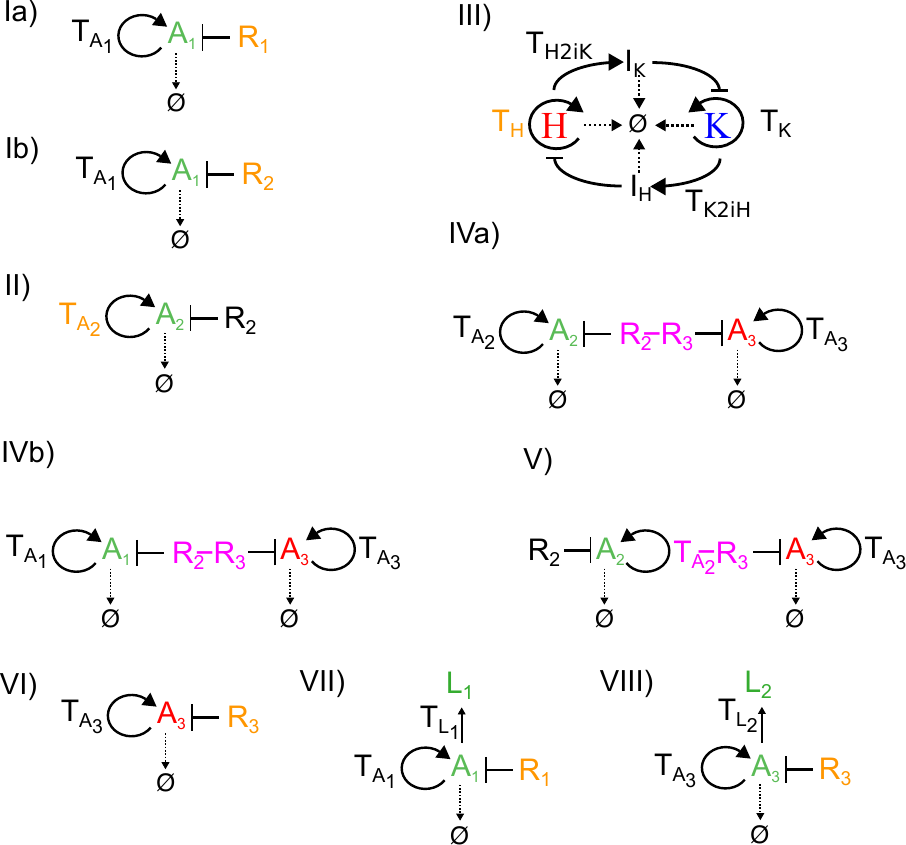}
        	\renewcommand{\figurename}{Supplementary Figure}\caption{Compendium of the reaction networks used in the Main Text. Each string of symbols stands for one DNA oligonucleotide sequence: H, K, or A$_i$ refer to autocatalytic nodes; R$_i$ are repressor templates; I$_H$ and I$_K$ are inhibitory species that hybridize on the template of the corresponding autocatalyst and slow down its production (\emph{26}); templates converting an input x into an output y are noted T$_{x2y}$ or simply T$_{x}$ when $y=x$ except for T$_{L_i}$ that converts the corresponding node into $L_i$; '-' indicates a hexaethyleneglycol spacer linking two different sequences. Normal arrows indicate \textit{activation} (\textit{self-activation} for circular arrows), arrows with blunt ends correspond to \textit{repression} (which can be \textit{inhibition} through a partial hybridization over the targeted template, or \textit{non-linear degradation} assisted by a repressor template). DNA strands used as bifurcation parameters (morphogens) in the experiments are noted in yellow and pink. Observed species are noted in green, red and blue.}
        	\label{fig:networks}
        \end{figure}

\begin{landscape}
        
     \begin{table}[H]
     	{\small\renewcommand{\tablename}{Supplementary Table}\caption{\label{tableSeqs} Sequences of oligonucleotides used in this work. Asterisks stand for phosphorothioate backbone modifications (3 or 4 phosphorothioates on the 5' end protect the strand from being degraded by the exonuclease). Nicking enzyme recognition sites are in bold, Nb.BsmI recognizes the site 5'-\textbf{GAATG$'$C}N-3', and Nt.BstNBI the site 5'-N$'$NNNN\textbf{GACTC}-3', where N is any nucleotide and $'$ indicates the position in the \emph{complementary} strand where a nick is created. The presence of a U in the nicking binding site of Nt.BstNBI, strongly reduces its binding affinity (\emph{26}). bt indicates a biotin, P indicates a phosphate. int.Spacer18 corresponds to a hexaethyleneglycol spacer. CY3.5, DY530, FAM and TexasRed are the used fluorophores.}}
	\vspace{-0.7cm}
    \footnotesize	     
     	\begin{center}
     		\begin{tabular}{|c | c |p{18cm}|}
     			\hline
     			Network & Name & Sequence $(5'-3')$\\
     			\hline
     			&  A$_1$\textsuperscript{$\dagger$} 	&\textbf{CATTC}AGGATC\\
			&  A$_2$\textsuperscript{$\dagger$}	&\textbf{CATTC}AGGATC\textbf{G}\\
     			& T$_{A_1}$	&CY3.5-C*G*A*T*CCT\textbf{GAATG$'$C}GATCCTGAA\\
     			&  R$_1$ &T*T*T*T*TCGATCCTGAATG-P\\
     			& R$_2$\textsuperscript{$\ddagger$} & bt-AAAAAACGATCCTGAATG-P\\
     	1-species  &  T$_{A_2}$\textsuperscript{$\ddagger$}& bt-*A*A*G*ATCCT\textbf{GAATG$'$C}GATCCTGAAT\\
     			bistable&   R$_2$-R$_3$\textsuperscript{$\ddagger$}& bt-AAAAAACGATCCTGAATG-int.Spacer18- T*T*T*T*CTCGTCAGAATG-P\\
     			& T$_{A_2}$-R$_3$\textsuperscript{$\ddagger$}& bt-*A*A*G*ATCCT\textbf{GAATG$'$C}GATCCTGAAT-int.Spacer18-T*T*T*T*CTCGTCAGAATG-P\\
     			& A$_3$	&\textbf{CATTC}TGACGA\textbf{G}\\
     			& T$_{A_3}$&DY530-*C*T*C*GTCA\textbf{GAATG$'$C}TCGTCAGAA\\
     			& R$_3$&T*T*T*T*CTCGTCAGAATG-P\\
     			\hline
     			& H	&CT\textbf{GAGTC}TTGG\\
     			& K	&TC\textbf{GAGTC}TGTT\\
     			& T$_{H}$\textsuperscript{$\mathsection$} & TexasRed-C*C*A*A\textbf{GACUC}AG$'$CCAA\textbf{GACTC}AGTTTTT-bt\\
     	2-species & T$_{K}$	&A*A*C*A\textbf{GACUC}GA$'$AACA\textbf{GACTC}GA-P\\
     	bistable    & I$_H$	&\textbf{GTC}TTGGCT\textbf{GAGT}AA\\
     			& I$_K$	&\textbf{GTC}TGTTTC\textbf{GAGT}AA\\
     			&  T$_{H2R_K}$&T*T*A*CTCGAAACAGAC$'$CCAA\textbf{GACTC}AG-FAM\\
     			& T$_{K2R_H}$	&T*T*A*CTCAGCCAAGAC$'$AACA\textbf{GACTC}GA-DY530\\
     			\hline     				        				    
     			& B$_1^r$ & G*G*A*TGAAGATGAGCATTACTTTCCGTCCCGAGAGACCTAACTGACACGCTTCCCATCGCTA-bt\\
     			  & B$_1^l$ & bt-AGCATTACTTTCCGTCCCGAGAGACCTAACTGACACGCTTCCCATCGCTAGGATGAAGATG-P\\
     			& T$_{L1}$&T*T*G*GATGAAGATGGGATGAAGATG\textbf{GAATG'C}GATCCTGAATG-P\\
     			 bead& L$_1$&CATCTTCATCCCATCTTCATCCAA\\
 & S & T*A*G*CGATGGGAAGCGTGTCAGTTAGGTCTCTCGGGACGGAAAGTAATGC-P\\    
      			sequences& B$_2^r$ & A*A*G*TAGGAGTAAGCATTACTTTCCGTCCCGAGAGACCTAACTGACACGCTTCCCATCGCTA-bt\\
      		 & B$_2^l$ & bt-AGCATTACTTTCCGTCCCGAGAGACCTAACTGACACGCTTCCCATCGCTAGATGTGGAGAG-P\\
           			& T$_{L2}$&T*T*G*ATGTGGAGAGAAGTAGGAGTA\textbf{GAATG'C}TCGTCAGAATG-P\\
      			& L$_2$&TACTCCTACTTCTCTCCACATCAA\\    			
     			\hline
     		\end{tabular}
     	\end{center}
     \end{table}
\vspace{-0.9cm}
{\footnotesize
     \textsuperscript{$\dagger$}  Species A$_{2}$ corresponds to the sequence of A$_{1}$ where one base on the 3' end has been removed. As a result, A$_1$ can initiate the autocatalysts templated by T$_{A_1}$ and T$_{A_2}$.
 
   \textsuperscript{$\ddagger$} bt in 5' + streptavidin also protects the corresponding strand from exonuclease degradation. In this case the last two bases in 5' (AA) are not replicated by the polymerase (see ref. \cite{Gines2017} for details). The presence of both bt and phosphorothioate in 5' is due to historical reasons.
     
     \textsuperscript{$\mathsection$} The presence of a TTTTT-bt on the 3' side allowed to attach this species to the surface through biotin-streptavidin (see Section \ref{sec:surface}). This modification was not needed for experiments in solution such as those presented in Figure 2 in the MT. However, for convenience, and because the modification does not play a role in the reaction network, the same sequence was used.}
     
    \end{landscape}
\normalsize      
      

   \subsection{Measurement of DNA concentrations}
   
     DNA concentration over space and time was measured by fluorescence. We used three different strategies:
    \begin{itemize}
    
     \item Recording the green fluorescence from EvaGreen dye (1x EvaGreen DNA binder,  20x dilution of the manufacturer's stock solution, Biotium). In this case, fluorescence is proportional to the concentration of double stranded DNA.
     
     \item Recording the fluorescence from dyes attached to the 3' or 5' end of template strands (see Supplementary Table~\ref{tableSeqs} for details). When the corresponding inputs or outputs hybridize on these templates, fluorescence is quenched.
    
    \item The concentrations of morphogens were measured with two different methods. For the 2-species bistable (Figure 2 of the MT) template T$_H$ was labeled with a Texas-Red fluorophore. For the 1-species bistable we added 1~$\mu$M of cascade blue-dextran M$_w = 3000$~Da  (Thermo Fisher Scientific, Molecular Probes) to the solution with high concentration of morphogen used to generate the gradient, and measured its fluorescence. The cascade blue-dextran has a molecular weight similar to the DNA templates and thus one expects a similar diffusion coefficient (Supplementary Figure~\ref{fig:gradFit}e-f). In a control experiment we compared the fluorescence profile obtained with cascade blue-dextran with a fluorescent DNA template, they were similar (Supplementary Figure~\ref{fig:gradFit}d). Note that in some experiments the gradient was imaged with ROX, a fluorophore that diffuses significantly faster than the morphogen. In those cases only the gradient at initial time is provided. In all other cases the gradient profile was recorded in real time. 
    
    The use of cascade-blue-dextran instead of a fluorescence-labeled oligonucleotide in 1-species experiments was due to constraints in the number of flurescence channels and the choice of fluorophores capable of N-quenching with the used sequences in the French flag experiments. The green channel was reserved for EvaGreen fluorescence. The yellow and red channels were used for DY530 and Cy3.5 dyes for monitoring species A$_1$ and A$_3$. Only the blue channel was left. Unfortunately, the signal of blue-fluorescent dyes attached to oligonucleotides was too low. We thus used cascade-blue-dextran at a much higher concentration than the morphogen to have enough signal.
	\end{itemize}
	
     In all kymographs and plots we represented the absolute value of the fluorescence shift, which is proportional to the concentration of fluorescent species in a concentration range between the detection limit and the concentration of the DNA strand complementary to the species of interest.
         
    \subsection{Temporal experiments}
Temporal experiments were performed on a BioRad CFX or Qiagen Rotor-Gene Q PCR machine (both used as temperature-controlled fluorimeters) using 20 $\mu$L of solution in 150 $\mu$L PCR tubes. Fluorophores ROX or Cy5-A$_{15}$ at 15 nM, where A$_{15}$ is an oligonucleotide with 15 adenines, were used as internal standards. The raw fluorescence intensity was divided by the intensity of the reference internal standard at each time point.

    \subsection{Spatio-temporal experiments}

Spatio-temporal experiments were performed within 50~mm$\times$4~mm$\times$0.2~mm glass capillaries (Vitrocom, USA). The capillary was loaded using a micropipette and a custom-made PDMS connector. 

\paragraph{Generation of the morphogen gradient.} The capillary was filled with 45~$\mu$L of the reaction solution without morphogen strand and then a pipette was inserted being in the 'push' position into the PDMS connector. The other end of the capillary was dipped into the reaction solution containing the morphogen DNA strand. 15 up-and-down pumps of 12.5~$\mu$L of solution were performed to create the gradient. Finally, the pipette was left in the 'pulled' position and the pipette together with the PDMS connector were removed.

\paragraph{Microscopy.}  Once the gradient was formed, the capillary was laid on a $5\times7.5$~cm glass slide.  5-minutes Araldite epoxy was used, both to seal the capillary ends and to glue them to the glass slide. No evaporation at all was observed for 48 h at 45°C. The fluorescence along the capillary was recorded on a Zeiss Axio Observer Z1 fully automated epifluorescence microscope equipped with a CoolLED pE-2 fiber-coupled illuminator, DAPI, GFP, YFP and RFP filter sets, a Marzhauser XY motorized stage, a Tokai Hit thermo plate, and Andor iXon Ultra 897 EMCCD camera, a shutter and a $2.5\times$ objective. These instruments were controlled with MicroManager 1.4. For optimal thermal conduction, mineral oil was added between the glass slide and the thermoplate on the microscope, and also between the glass slide and the capillary. For each capillary, 16 contiguous 3.17$\times$3.17 mm$^2$ ($128\times128$ pixel$^2$) images were recorded automatically every 1 to 10 minutes.  Multi-color fluorescence microscopy was used to record the concentration of different DNA species over time. Images of the beads were acquired in bright field with a 10 or a $40\times$ objective.

\paragraph{Image treatment.} The raw data were treated with ImageJ / Fiji (NIH) and Matlab (The Mathworks). 
Prior to data analysis, the 16 images were stitched together without overlapping. Subsequently, the inhomogeneous illumination was corrected by two different protocols. When the initial concentration was flat (for all species except for the morphogen) a division by the first frame was performed. When the initial concentration was not homogeneous, for example for the morphogen, a polynome was fitted to the raw data of one of the 3.17$\times$3.17 mm$^2$ of the first frame and each image was divided by it.

The counting of the particles (aggregates or single bead without distinction) has been done on images acquired in bright field with a 10 or a $40\times$ objective. Contiguous images were stitched together without overlapping. Images were first binarized. Then, the Fiji Particle Analysis plugins was used to determine the number of particles in contiguous areas (100$\mu$m to 250$\mu$m wide) along the capillary.

  \subsection{Data treatment}

        \paragraph{Kymographs.} Considering the 50~mm $\times 4$~mm capillary as a 1-dimensional reactor, we first averaged the corrected fluorescence images over the width of the capillary (along the $y$ axis). The kymographs were obtained by stacking these profiles over time.

        \paragraph{Front position and width.}  The profiles averaged along $y$ were further 
averaged along the $x$ axis by performing a moving average over 25 pixels and subsequently normalized between 0 and 1. A sigmoidal function $f(x)=\frac{1}{1+e^{(-x+x_0)/\lambda}}$ was fitted to these data. The front position corresponds to $x_0$ while the width of the front is defined as $\lambda$. When the sigmoidal fit failed, the position of the fronts was determined by finding the $x$ coordinate for which the value of the normalized profile was closest to 0.5. 
               
   %

        
     	 \paragraph{Front velocity.} The data of front position over time was fitted by a polynomial function to reduce noise. The velocity was calculated as the time derivative of this polynomial fit.
	 	       
	    \paragraph{Morphogen concentration at the front position.} The morphogen concentration was either obtained directly by measuring the fluorescence of species T$_H$ (Figure 2 of the Main Text) or indirectly by measuring the fluorescence of the cascade blue-dextran gradient (Figures 1 and 3 of the Main Text). In all cases, two control channels containing a 0 and constant concentration of the fluorescent species were used in each experiment to calibrate the intensity to concentration response. The fluorescence signal of the morphogen was shifted to 0 where the concentration of the morphogen was expected to be 0~nM (i. e. at $x=0$) and then converted to concentration by using the intensity to concentration relation previously determined.

\newpage

\section{Data associated to Figure 1 in the Main Text}

\subsection{The 1-species bistable in well-mixed conditions}

          \begin{figure}[H]
          	\centering
	\subfloat[]{  
          	\includegraphics[width=0.7\linewidth]{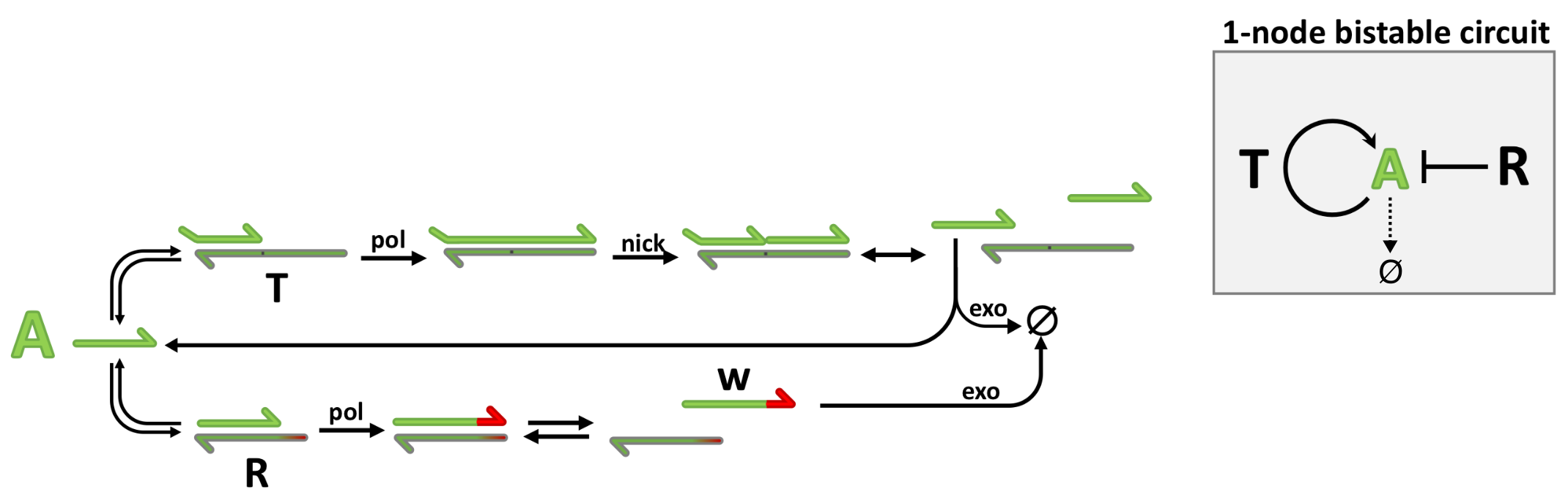}
	}
	
	\subfloat[]{  
	\includegraphics[width=16cm]{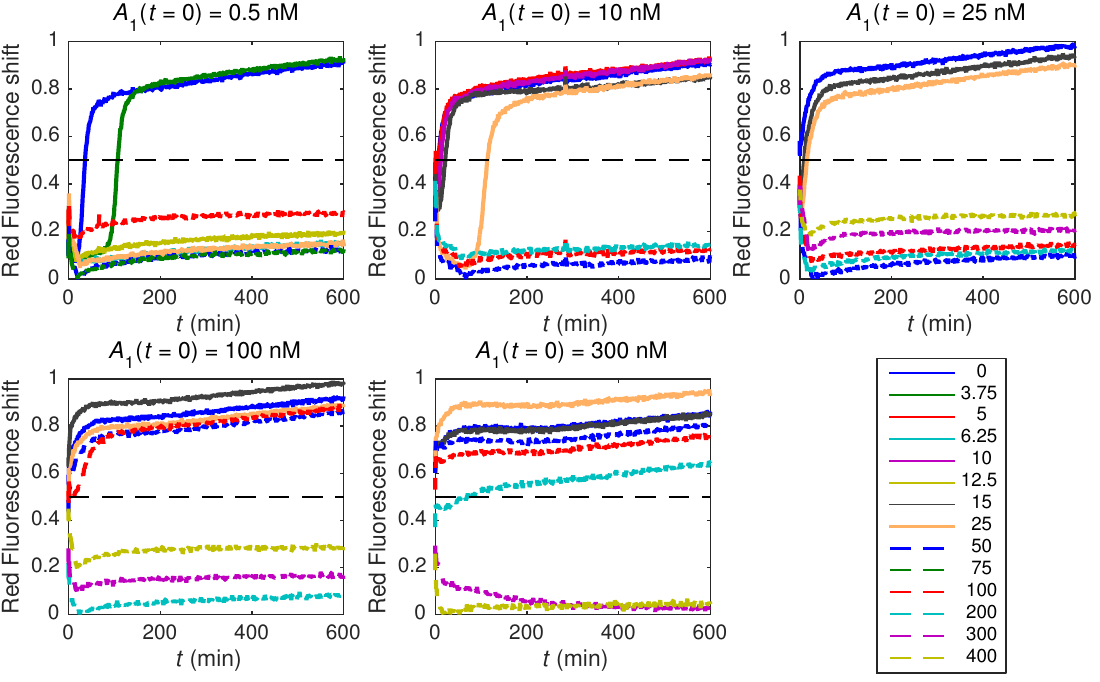}
	}
          	\renewcommand{\figurename}{Supplementary Figure}\caption{Detailed mechanism (a) and well-mixed dynamics (b) of the 1-species bistable circuit. (a) The inset reminds the topology of the network. This circuit uses a single autocatalytic template T producing A exponentially, coupled to a species-specific degradation mechanism driven by the repressor R. Only species that are not protected on the 5' end are degraded by the exonuclease (the top strands). This molecular program admits 2 stable states, defined by the amplification and non-amplification of the A strand. (b) Dynamics of the 1-species bistable for different initial conditions and different thresholded degradation rates. Red fluorescence shift coming from Cy3.5 dye \emph{vs} time. Each plot represents a different initial condition of A$_1$ as indicated in the title and each curve represents a different thresholded degradation rate given by the concentration of species R$_1$ (see legend, in nM). When the final fluorescence shift was above the dashed black horizontal line the bistable was considered to be in state 1 (high) and otherwise in state 0 (low). $T=45^\circ$C, $T_{A_1} = 25$~nM.\label{fig:MontBif1}}
\end{figure}

        \begin{figure}[H]
       	\centering	  
       \includegraphics[height=6cm]{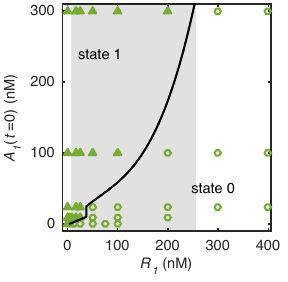} 
       	\renewcommand{\figurename}{Supplementary Figure}\caption{Phase space of the 1-species bistable network as a function of repressor concentration, $R_1$, and initial condition $A_1(t=0)$. Triangles and circles correspond to systems ending in the high and low concentration state, respectively. The shaded region indicates the range of $R_1$ where the system is bistable. Plotted from data in Supplementary Figure~\ref{fig:MontBif1}.}
       	\label{fig:MontBif3}
       \end{figure}

\subsection{Characterization of the morphogen gradient and its variation over time}
 
%

   During the 10 to 40 hours-long experiments the morphogen gradient diffuses slightly. Here, we estimate an order of magnitude of the variation of the gradient over time. We have
   
   \begin{equation}
   \partial_t M(x,t) = D \partial_{xx} M(x,t)
   \end{equation}
   
\noindent where $M$ is the concentration of morphogen, $D$ its diffusion coefficient, $x$ the spatial coordinate along the channel and $t$ the time. To get an upper limit to the change of $M(x,t)$, we take the second derivative at $t =0$,
   
   \begin{equation}
   \partial_t M(x,t) \leq D \partial_{xx} M(x, 0).
   \end{equation}
   If $M(x, 0) = M_{max}e^{(x-x_0)/l}$, as we show in Supplementary Figure \ref{fig:gradFit}b, we have 
   
   \begin{equation}
   \partial_t M(x,t) \leq \frac{D}{l^2} M(x, 0).
   \end{equation}
   We can now integrate the differential equation over time $\Delta t$ and we get
   
   \begin{equation}
   \frac{\Delta M(x)}{M(x,0)} \leq \frac{D\Delta t}{l^2},
   \end{equation}
   where $\Delta M(x) = M(x,t) - M(x,0)$. Taking $D = 10^{-2}$~mm$^2$/min \cite{Zadorin2015}, $l \approx 10$~mm and $\Delta t = 10^3$ min (16~h) we have
   
   \begin{equation}
   \frac{\Delta M(x)}{M(x,0)} \leq 0.1.
   \label{eq:morphoDrift}
   \end{equation}
   In conclusion, the maximum concentration change over 16~h is lower than 10\%.
   
   To test this estimation experimentally, Supplementary Figure \ref{fig:gradFit}f shows the profile of a gradient of fluorescein-labeled species R$_1$ recorded for 50~h. Over 50 h the spatial drift  in the middle of the channel was 2~mm and  the concentration drift was 10 \%. The maximum concentration drift (observed on the right side of the capillary) was 30\%, in agreement with Equation~\ref{eq:morphoDrift}.
   
   \begin{figure}[H]
                 
          	\centering
		\subfloat[]{\includegraphics[width=0.3\linewidth]{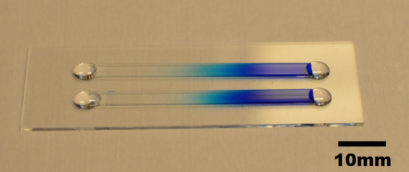}}
		\hspace{0.5cm}
\subfloat[]{\includegraphics[width=0.3\linewidth]{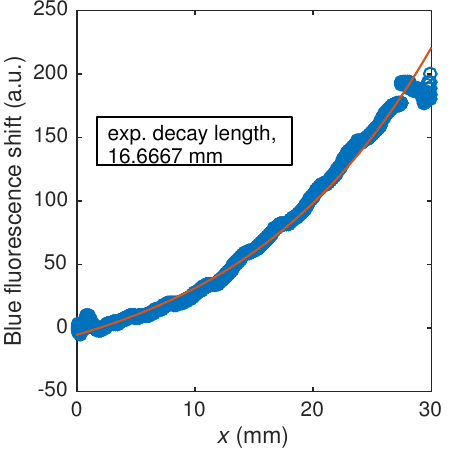}}	
\hspace{0.5cm}	
		\subfloat[]{\includegraphics[width=0.3\linewidth]{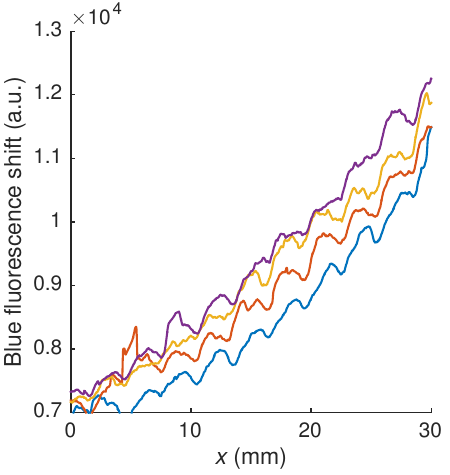}}
		
		\subfloat[]{\includegraphics[width=0.3\linewidth]{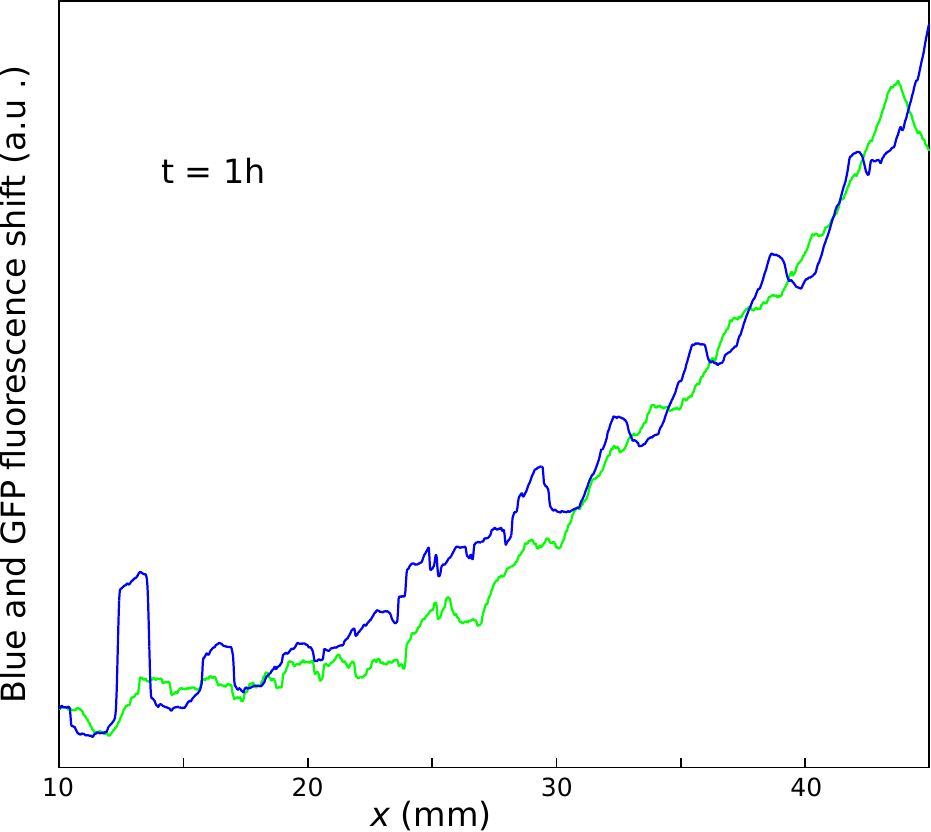}}		
		\hspace{0.5cm}
		\subfloat[]{\includegraphics[width=0.3\linewidth]{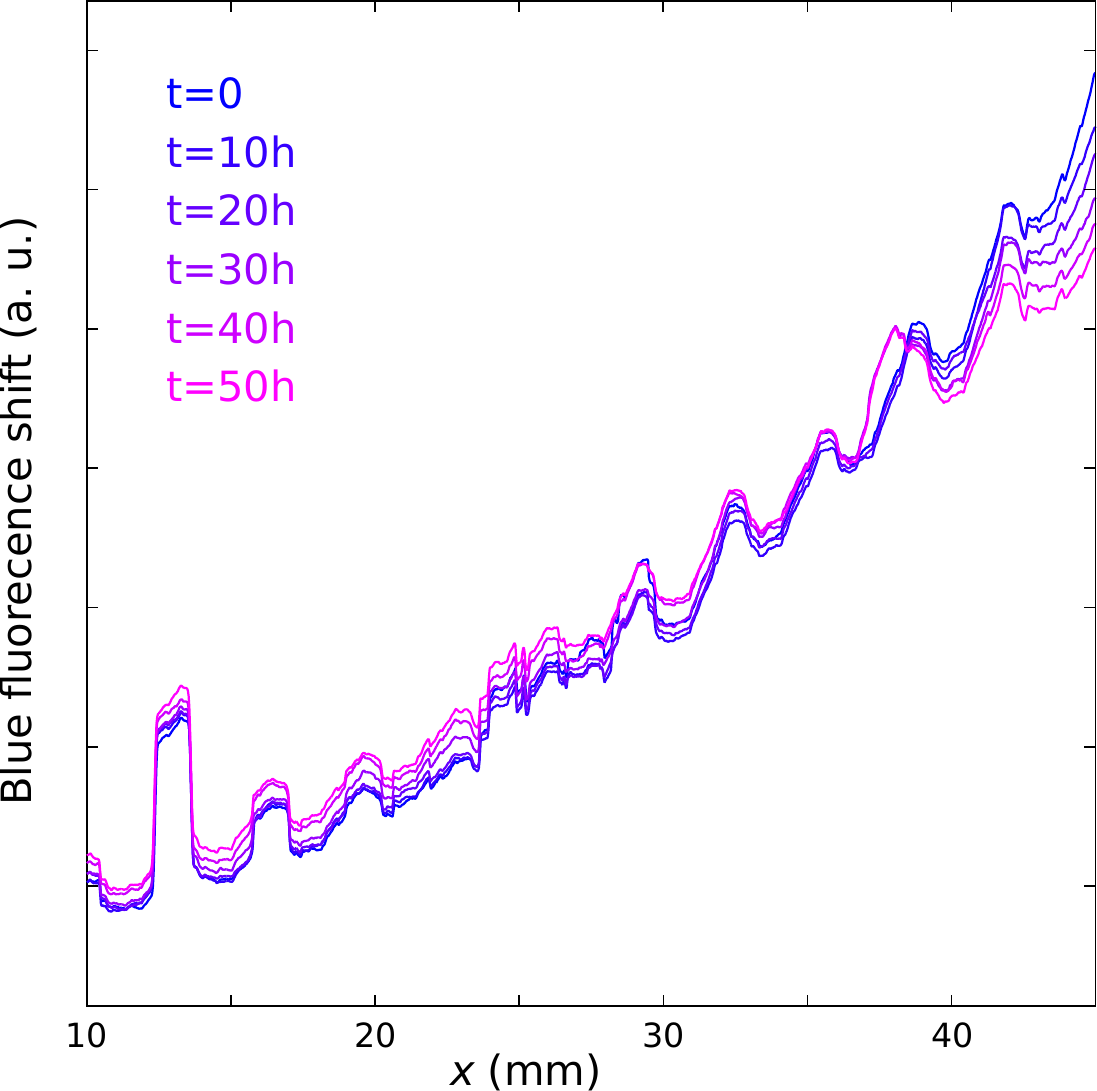}}
		\hspace{0.5cm}
		\subfloat[]{\includegraphics[width=0.3\linewidth]{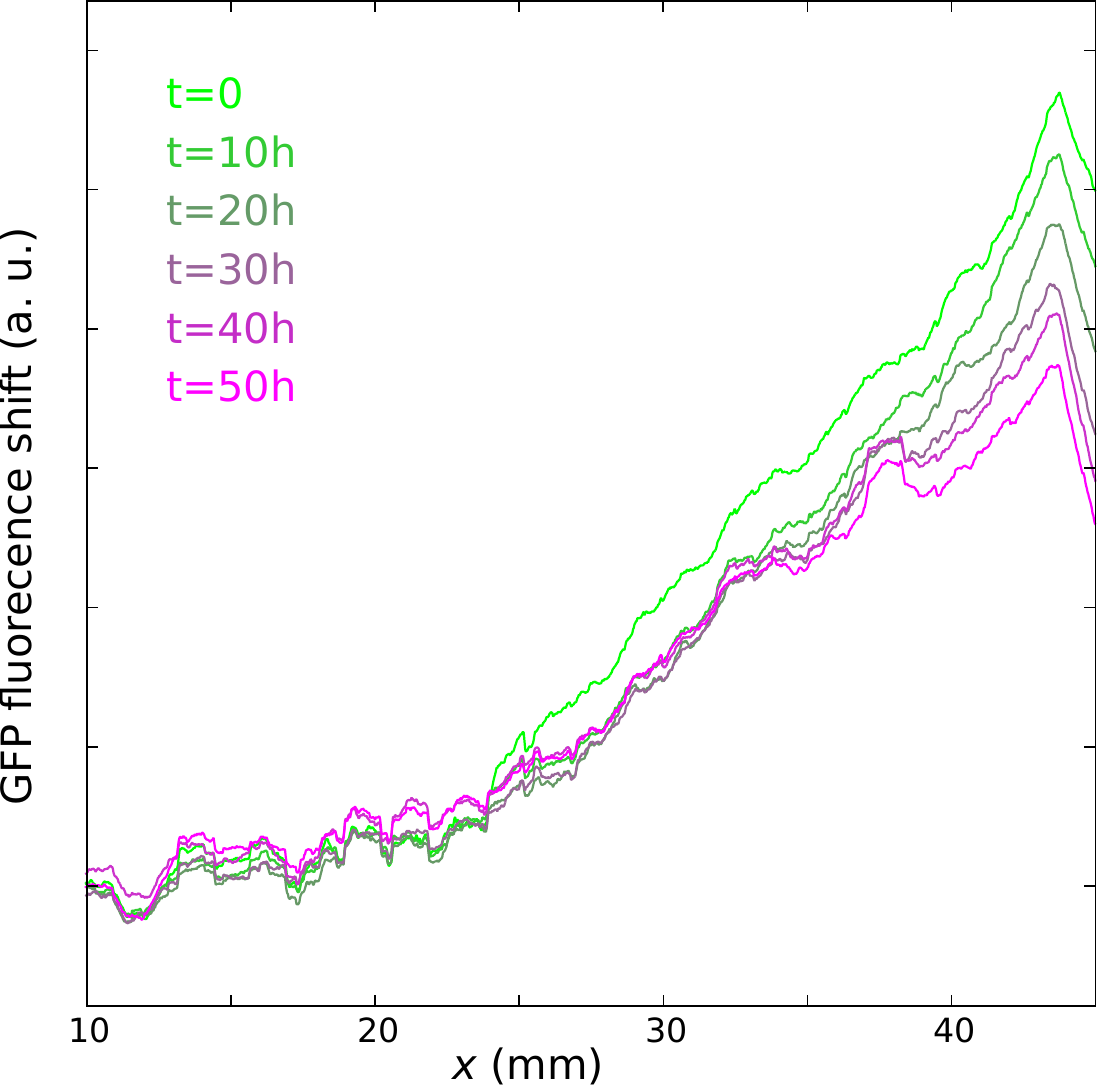}}

          	\renewcommand{\figurename}{Supplementary Figure}\caption{Characterization of the morphogen gradient. (a) Gradients of methylene blue dye generated inside the glass capillaries by pipetting back and forth a constant volume of dye. Each extremity of the capillary is sealed with a droplet of glue that serves also to attach it to a glass slide. Methylene blue is not present in the experiments and is used here solely for visualization.
	(b) The characteristic length of the morphogen gradient is 17~mm. Exponential fit (red line) to the fluorescence intensity from the cascade-blue-dextran gradient in Figure~\ref{fig1} of the Main Text.
	(c) The generation of the morphogen gradient is fairly reproducible. Fluorescence profiles of cascade-blue-dextran gradients in four different channels prepared at the same time. The gradients were slide-averaged over 25 pixels. The sawtooth shape of the curves is due to  illumination inhomogeneities that were not corrected here.
	(d) The initial gradient of cascade-blue-dextran (blue line) is similar to the one of fluorescein labeled R$_1$ (green line). The dynamics of the gradients of cascade-blue-dextran (e) and fluorescein labeled R$_1$ (f) at 45 °C are not identical but fairly similar. Each curve represents the fluorescence profile every 10~h.  (d-f) display results from a single capillary where the gradients of cascade-blue-dextran and fluorescein labeled R$_1$ were generated simultaneously.
	\label{fig:gradFit}}
          \end{figure}

  \begin{figure}[H]
 	\centering 
	\subfloat[]{\includegraphics[width=8cm]{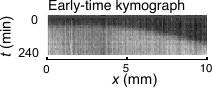}}
  
 	\subfloat[]{\includegraphics[width=0.7\linewidth]{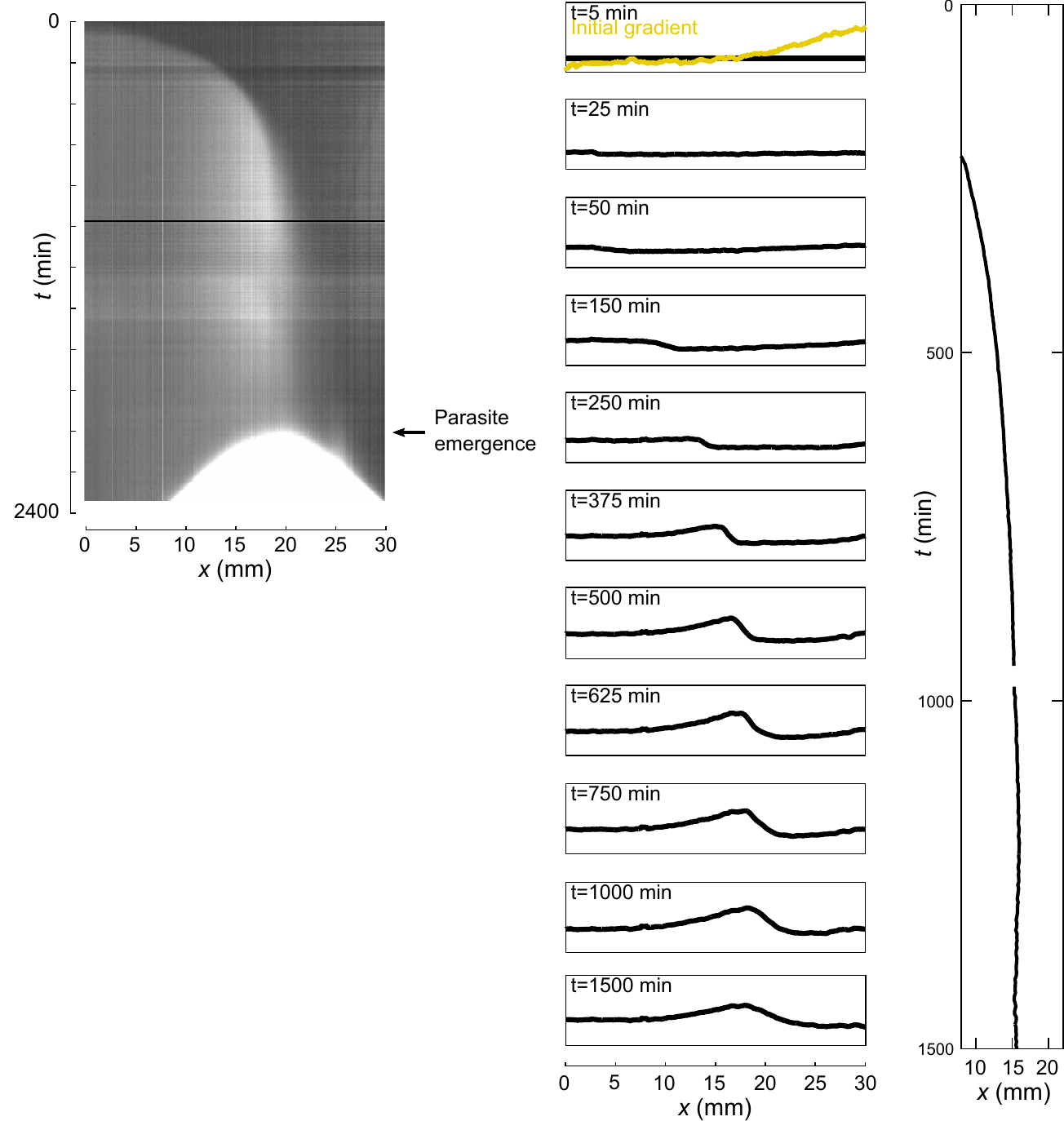}}
 	\renewcommand{\figurename}{Supplementary Figure}\caption{Short time and long time behavior of the Polish flag pattern. (a) Kymograph of Figure 1 at short times. At $t\approx 70$~min a purely reaction mechanism creates a front at $x \approx 5$~mm. Later this front moves slowly through a reaction-diffusion mechanism from left to right. 
	(b) Polish flag patterning lasts up to 30 h. Experiment similar to  the one shown in Figure 1D-E in the Main Text. Each panel represents the kymograph (left), the fluorescence profiles along the channel at different times (center) with EvaGreen fluorescence in black and morphogen fluorescence in yellow (this last one only at initial time), and the position of the front (right). The fluorescence of EvaGreen is proportional to the concentration of dsDNA in the capillary. At $2000$~min a strongly fluorescent species, called a 'parasite', emerges and kills the immobile front (see Supplementary Figure \ref{fig:parasite} for a discussion).  Network Ib with R$_2=0-400$~nM, and initial constant concentrations of $T_{A_1}=25$~nM and $A_1 = 0.5$~nM. $T = 45^{\circ}$C. }
 	\label{fig:1Polish_flag_pT}
 \end{figure}

\subsection{Data on the emergence of the parasite}
 
In Supplementary Figure \ref{fig:1Polish_flag_pT} a very high increase of fluorescence appears on the kymograph after 2000 min of experiment. We call the species, or mixture of species, associated to this high fluorescence a 'parasite'. Such a 'parasite' is intrinsic to the isothermal exponential amplification reaction scheme (EXPAR) ---that couples a polymerase and a nicking enzyme and that is at the core of the PEN DNA toolbox--- and it  corresponds to a late-phase non-specific amplification \cite{Tan2008}. Indeed, even in the absence of an autocatalytic template, the combination of a polymerase, a nicking enzyme and  dNTPs, results in the emergence of one or several autocatalytic DNA sequences that are highly fluorescent. Sequencing of parasite species has revealed that these species are mainly poly(AT) sequences with occasional T/G and A/C substitutions, and occasional palyndromic stretches containing the nicking enzyme recognition sequence \cite{Tan2008}.

\begin{figure}[H]
 	\centering  
 	\includegraphics[width=\linewidth]{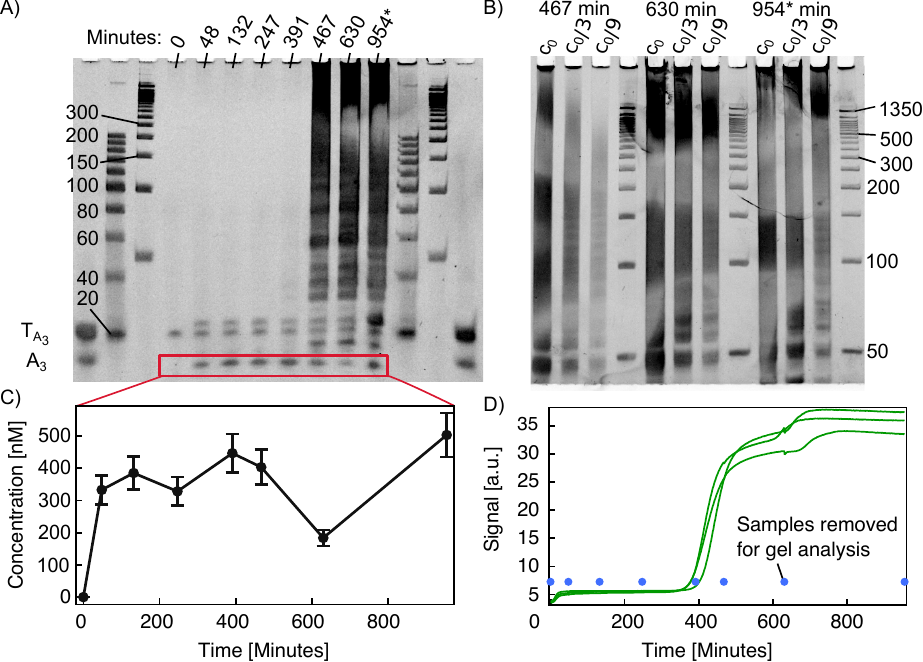}
 	\renewcommand{\figurename}{Supplementary Figure}\caption{Characterization of the parasite that emerges in the autocatalytic reaction. A) The gel shows the reaction products after different incubation times at 45°C. The sample labeled with * was left at room temperature for 6 hours after that. The designed product of the reaction, A$_3$, is marked with a red rectangle and used for further analysis in C). After 400 minutes, the parasite becomes clearly visible. B) To better separate the longer products observed after 400 minutes, a second gel was used with an increased run-time. The samples were added in three different concentrations. C) The concentration of the autocatalytic replicator obtained from the gel in A) is plotted against time. The concentration reaches a plateau phase after 50 minutes, but shows changes when the parasite appears after 400 minutes. D) EvaGreen real-time fluorescence signal of the same experiment, in triplicates. The designed autocatalysis occurs within the first 25 min, the jump at 400 min is due to the parasite. The blue dots indicate times at which samples were removed for gel analysis. The fluorescent signal is in good agreement with the gel in panel A. Conditions: $A_3 = 1$~nM, $T_{A_3}=50$~nM, $pol = 0.3\%$, $nick = 1\%$, $exo = 1\%$.}
 	\label{fig:parasite}
 \end{figure}

\begin{figure}[H]
 	\centering  
	\includegraphics[width=\linewidth]{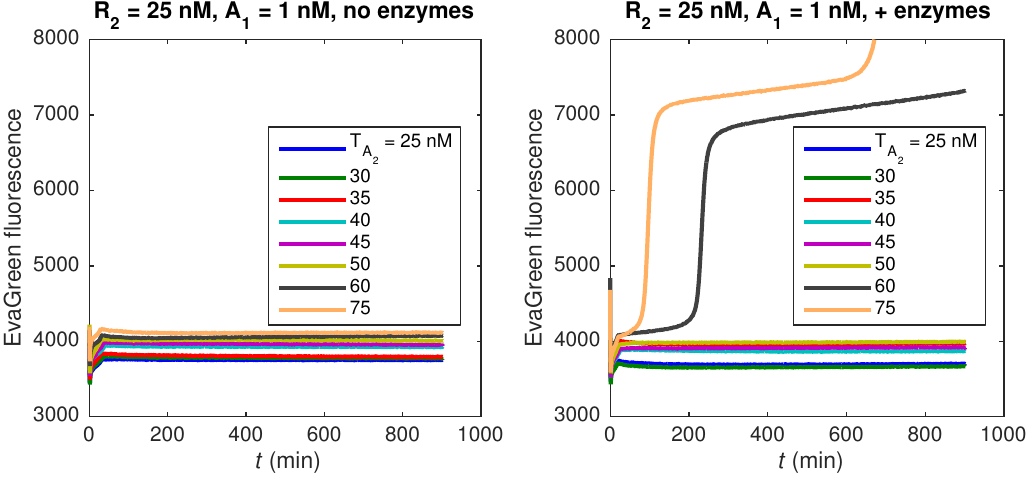}
 	\renewcommand{\figurename}{Supplementary Figure}\caption{Increasing $T_{A_2}$ makes the network II switch from state low to state high. EvaGreen fluorescence versus time in the absence (left) and in the presence of enzymes with $A_1 =10$~nM and different $T_{A_2}$ (colored lines). $T = 45^{\circ}$C. }
 	\label{fig:1Bifurc_CB}
 \end{figure}

\begin{figure}[H]
 	\centering  
 	\includegraphics[scale=0.8]{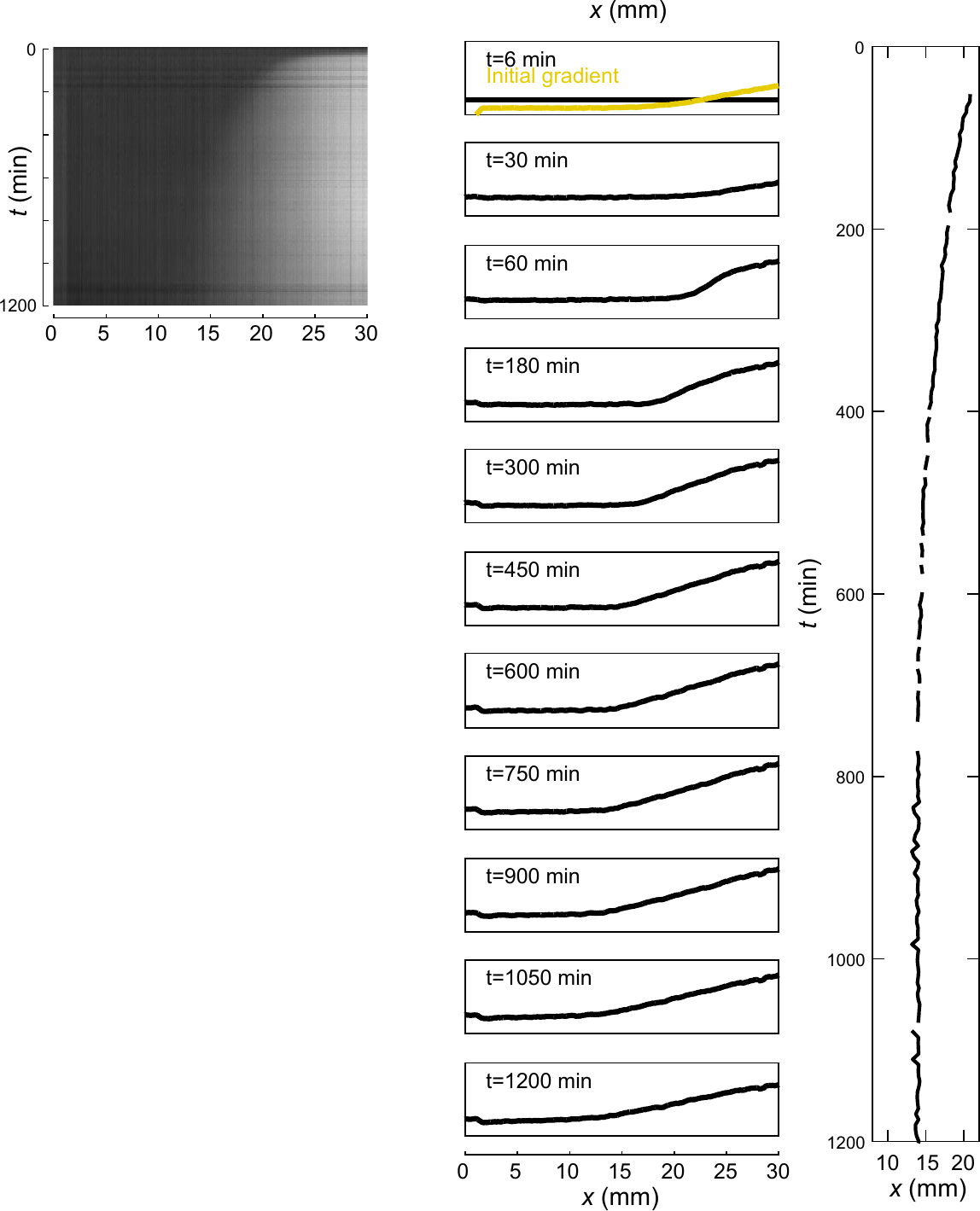}
 	\renewcommand{\figurename}{Supplementary Figure}\caption{Complementary Polish flag pattern using the autocatalytic template instead of the repressor as morphogen. Each panel represents the kymograph (left), the fluorescence profiles along the channel at different times (center) with EvaGreen fluorescence in black and morphogen fluorescence in yellow (this last one only at initial time), and the position of the front (right). Network II with a gradient of T$_{A_2}$ from 25 to 100~nM, and initial constant concentrations of  $R_2=25$~nM, $A_2 = 10$~nM.  $T = 45^{\circ}$C. }
 	\label{fig:1Polish_flag_CB}
 \end{figure}

\begin{figure}[H]
        	\centering  
	\includegraphics[width=0.4\linewidth]{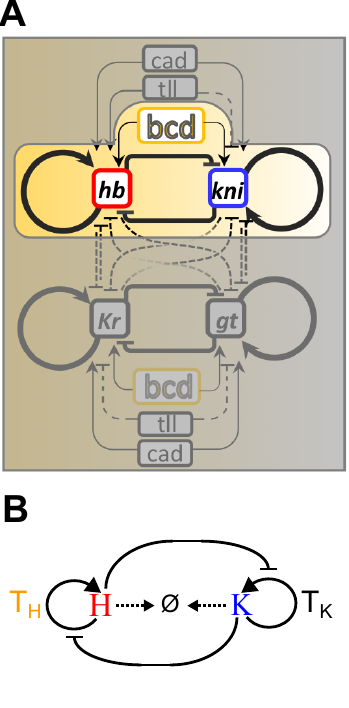}
  	\renewcommand{\figurename}{Supplementary Figure}\caption{(\textbf{A}) Topology of the gap gene network, from \cite{Manu2009}, the portion emulated here is highlighted. (\textbf{B}) Analogue DNA-based network for the principal interactions between Bicoid, bcd, Hunchback, hb and Knirps, kni. Their DNA couterparts are respectively noted T$_{H}$, H and K. 
	\label{fig:BicoidNetwork}}
        \end{figure}

We used denaturing PAGE-analysis (12.5\%-PAA gel, 19:1, 50\% urea, TBE-buffer, SYBR Gold staining) to follow the time-evolution of the amplification reaction of A$_3$ in the presence of T$_3$ (Supplementary Figure \ref{fig:parasite}). 

 \subsection{The autocatalyst template also acts as a bifurcation parameter}

\section{Data associated to Figure 2 in the Main Text}

      \begin{figure}[H]
        	\centering  
	\subfloat[]{\includegraphics[width=\linewidth]{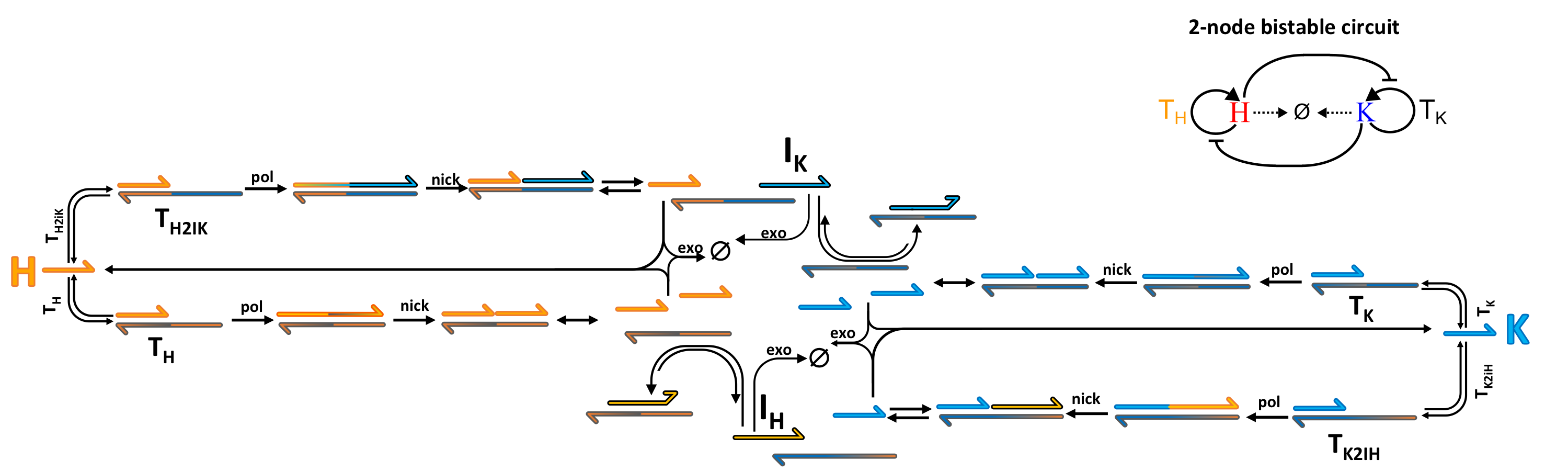}}

        \subfloat[]{\includegraphics[width=0.4\linewidth]{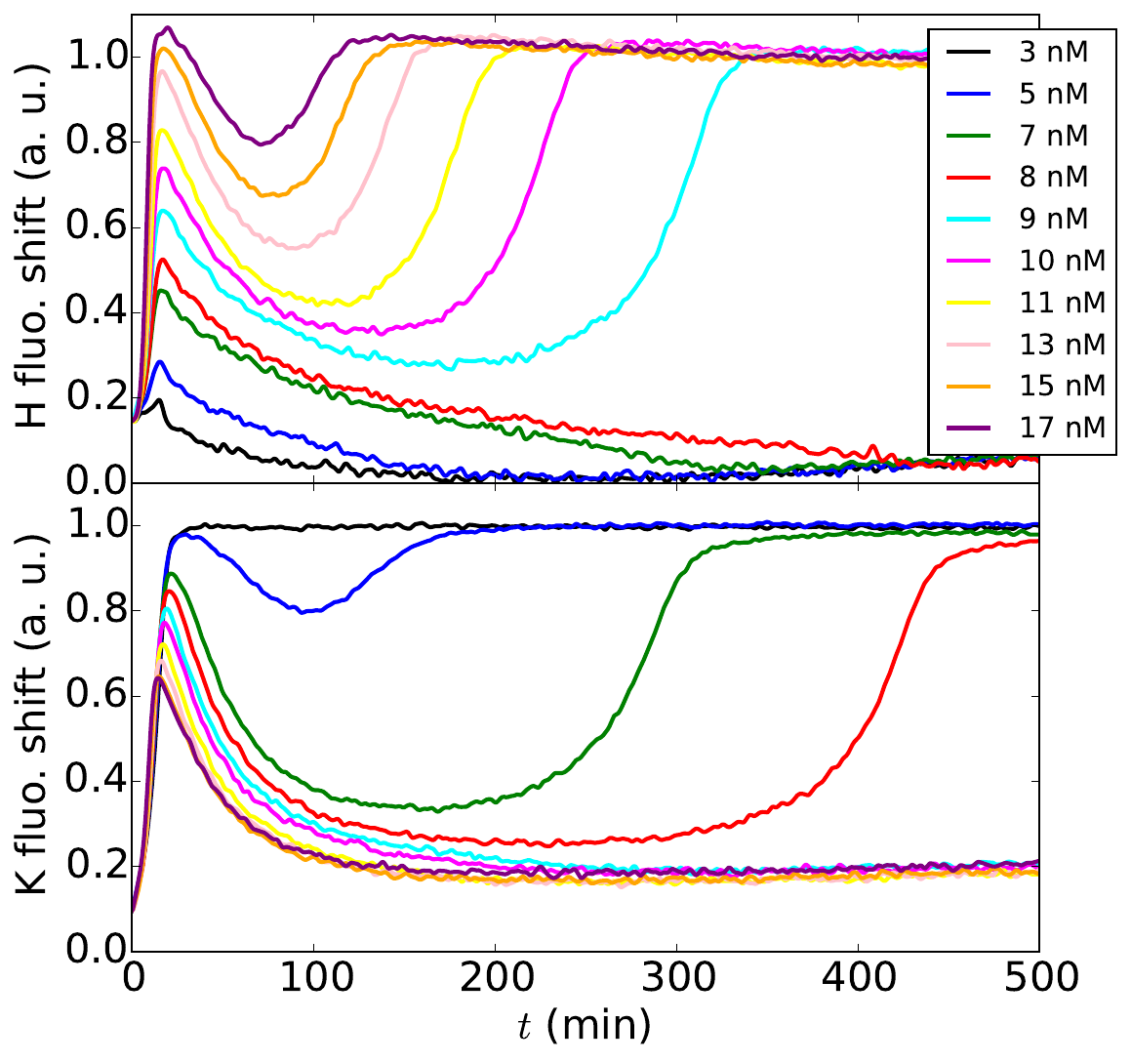}}
	\hspace{1cm}
	\subfloat[]{\includegraphics[width=0.4\linewidth]{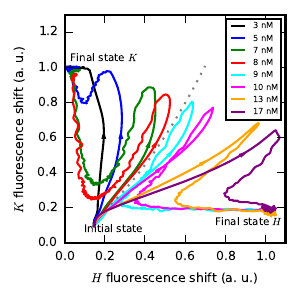}}
	
        	\renewcommand{\figurename}{Supplementary Figure}\caption{Detailed mechanism (a) and well-mixed dynamics (b, c) of the 2-species bistable circuit.
	 (a) The inset reminds the simplified topology. This circuit uses 4 catalytic templates: T$_{H}$ and T$_{K}$ catalyze the exponential amplification of H and K respectively. The templates T$_{H2I_K}$  and  T$_{K2I_H}$ initiate the synthesis of I$_K$ and I$_H$, using respectively H and K strands as input. I$_K$ and I$_H$ are able to bind the cognate autocatalytic template with a higher affinity that their input, preventing therefore the amplification of the H and K nodes. Only species that are not protected on the 5' end are degraded by the exonuclease (the top strands). By mixing these four templates, a 2-species bistable circuit is obtained, leading to two stable states where either H or K are amplified, depending on the experimental initial conditions.
	(b) The concentration of T$_H$ is a bifurcation parameter for the network in panel a. Fluorescence shift due to species H (top) and K (bottom) as a function of time in well-mixed conditions. Curves with same colors refer to identical experiments, each with a different concentration of template T$_{H}$ that controls the growth rate of H. When the concentration of T$_{H}$ is 8 nM or less, K grows and H dies and the final state is $(H,K) = (0, 1)$. When the concentration of T$_{H}$ is 9 nM or more, K dies and H grows and the final state is $(H,K) = (1, 0)$. $T=42^\circ$, $T_{K}=20$~nM,  $T_{H2R_K}=20$~nM, $T_{K2R_H}=20$~nM, initial conditions $H=K=0.5$~nM.
	(c) Phase portrait from data in panel b. The dashed line separates the $T_H$ values for which the final state is $K$ and those for which the final state is $H$.	\label{fig:PadiracBistable}}
        \end{figure}


\subsection{Spatio-temporal simulations in Figure 2 \label{sec:simu}}
    
    Following \cite{Padirac2012} the 1-dimensional spatio-temporal dynamics of the 2-species network were simulated with a four variable model for species H, K, I$_H$ and $I_K$, the first two being the autocatalysts and the last two, respectively, the inhibitors of H and K.

\begin{eqnarray}
\frac{\partial H}{\partial t}  &=& M(x) \frac{k_p^H H}{1+\lambda H +\lambda_I I_H} - k_{deg}H + D \frac{\partial^2 H}{\partial x^2} + \epsilon \nonumber\\
\frac{\partial K}{\partial t}  &=& \frac{k_p^K K}{1+\lambda K +\lambda_I I_K} - k_{deg} K + D \frac{\partial^2 K}{\partial x^2} + \epsilon \label{simuModel}\\
\frac{\partial I_H}{\partial t}  &=& \frac{k^{I_H}_p K}{1+\lambda K} - k^I_{deg}I_H + D \frac{\partial^2 I_H}{\partial x^2} \nonumber\\
\frac{\partial I_K}{\partial t}  &=& \frac{k^{I_K}_p H}{1+\lambda H} - k^I_{deg}I_K + D \frac{\partial^2 I_K}{\partial x^2} \nonumber
\end{eqnarray}
    with initial conditions
    \begin{equation}
    H(x,0) = H_0, \,\,\,  K(x,0) = K_0,  \,\,\,  I_K(x,0) = I_K(x,0) = 0, \label{simuIC}
    \end{equation}
and null Neumann boundary conditions.

$M(x)=30e^{-(x-20)/6}+0.1$ is the morphogen gradient, $k_p$ is the rate constant for production, $k_{deg}$ and $k^i_{deg}$ the  degradation rate constants for autocatalysts and inhibitors, respectively, $\lambda$ and $\lambda_I$ the factors accounting for the inhibition of the species by its activator and its inhibitor, respectively. Although mathematically similar, these two terms have different chemical origin. An excess of activator (e.g. K) for a given species (e.g. I$_H$) increases the hybridization on the corresponding template strand, which is the substrate of the polymerase and saturates it. An excess of inhibitor (e.g.  I$_K$) for a given autocatalytic species (e.g. K) increases its hybridization on the corresponding template and prevents the hybridization of the activation species (see Supplementary Figure \ref{fig:PadiracBistable} for more details on the mechanism). $D$ is the diffusion coefficient for all species. $\epsilon$ is a constant accounting for the leak reaction that generates an autocatalytic species in the absence of its activator.

Equations (\ref{simuModel}--\ref{simuIC}) with the parameters detailed in Supplementary Table~\ref{tab:simuParam} were solved in Mathematica using the method of lines, space was discretized with a minimum of 400 points. 

\begin{table}[h!]
\renewcommand{\figurename}{Supplementary Table}\caption{Numerical values of the parameters used to numerically solve Equations (\ref{simuModel}--\ref{simuIC}). cau means concentration arbitrary units. All the parameters were extracted from a manual fit to a series of independent well-mixed experiments except for $D$ that was measured in ref. \cite{Zadorin2015}.  \label{tab:simuParam}}
\begin{center}
\begin{tabular}{lr}
\hline
Parameter & Value  \\
\hline
$k_p^H$		&	0.012~nM$^{-1}$min$^{-1}$\\
$k_p^K$		&	0.15~min$^{-1}$\\
$k_p^{I_H}$	&	0.005~min$^{-1}$\\
$k_p^{I_K}$	&	0.02~min$^{-1}$\\
$k_{deg}$  	& 	0.05~min$^{-1}$\\
$k^I_{deg}$	&	0.005~min$^{-1}$\\
$\lambda$ 	&	0.01~cau$^{-1}$\\
$\lambda_I$	&	10~cau$^{-1}$\\	
$D$			&	0.018~mm$^2$/min\\
 $\epsilon$ 	&	$10^{-7}$~cau/min\\
  $H_0$		& 	$10^{-3}$~cau\\
 $K_0$		&	$10^{-3}$~cau\\
 Integration time &	900~min\\
 Integration distance &	30~mm \\
\hline
\end{tabular}
\end{center}
\end{table}

To determine the parameters in Supplementary Table~\ref{tab:simuParam} we fitted Equation~\ref{simuModel} with $D=0$ to a series of independent experiments in well-mixed conditions (Supplementary Figure~\ref{fig:PadiracBistable}b) where the concentration of T$_H$ (i.e. $M$ in Equation~\ref{simuModel}) changed. The results of the fit are shown in Supplementary Figure~\ref{fig:fitPadiracData}. Due to the simplicity of the model only some features of the dynamics for $t<100$~min are captured. Notably the T$_H$ concentration value at which bifurcation occurs and the non-monotonous kinetics close to this bifurcation point at $T_H=7-9$~nM. The simple model is in very good agreement with the data for $t>100$~min.

  \begin{figure}[H]
     	\centering  
     	\includegraphics[width=\linewidth]{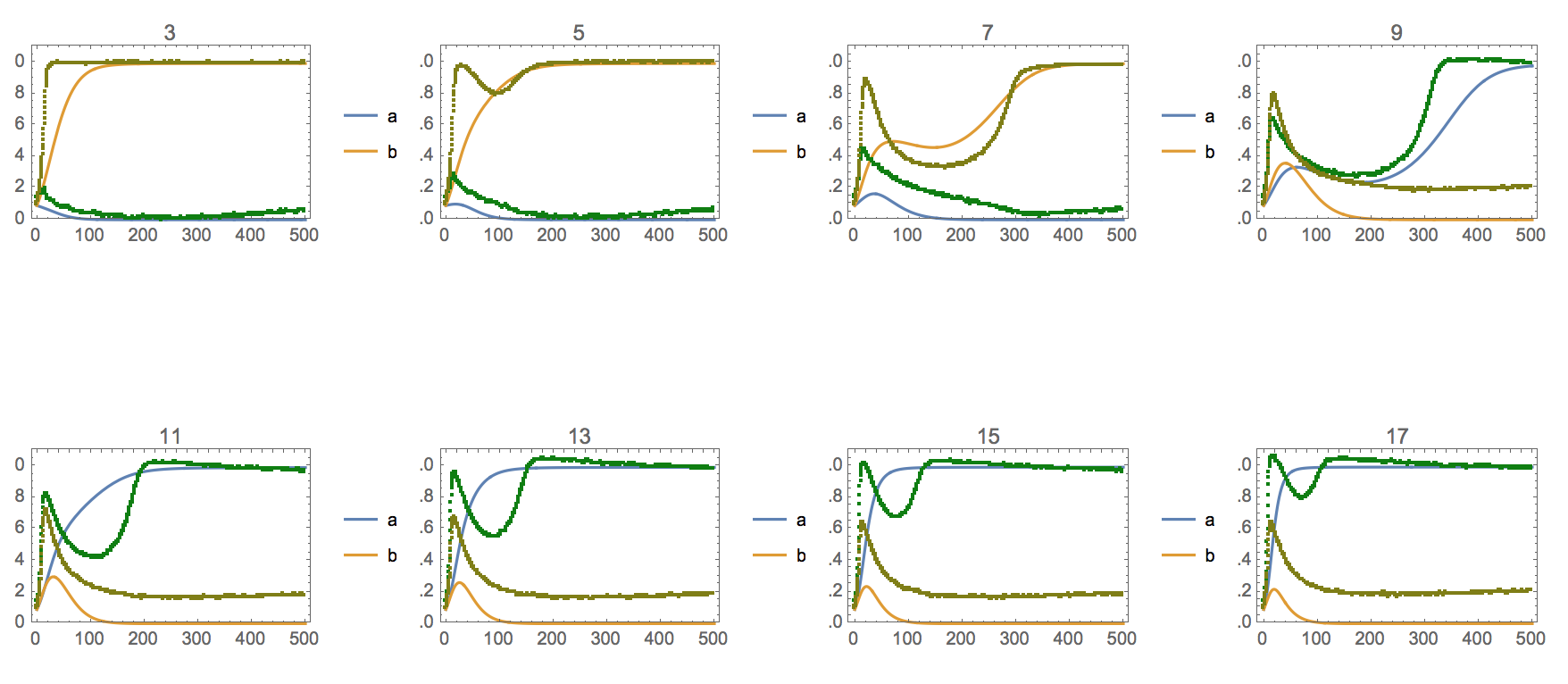}
     	\renewcommand{\figurename}{Supplementary Figure}\caption{Independent determination of the kinetic rates of the 2-species bistable in well-mixed conditions. Each plot represents fluorescence intensity vs. time (in min) for different concentrations of T$_H$ in nM (number above the plots). The brown and green dots represent the data in Supplementary Figure~\ref{fig:PadiracBistable}b (see caption for details). The blue and yellow lines are the manual fits of Equation~\ref{simuModel} with $D=0$. The obtained parameters are in Supplementary Table~\ref{tab:simuParam}.}
     	\label{fig:fitPadiracData}
     \end{figure}

    \subsection{Reproducibility of the pattern generated by the 2-species bistable network}

\begin{figure}[H]
     	\centering  
     	\includegraphics[width=0.8\linewidth]{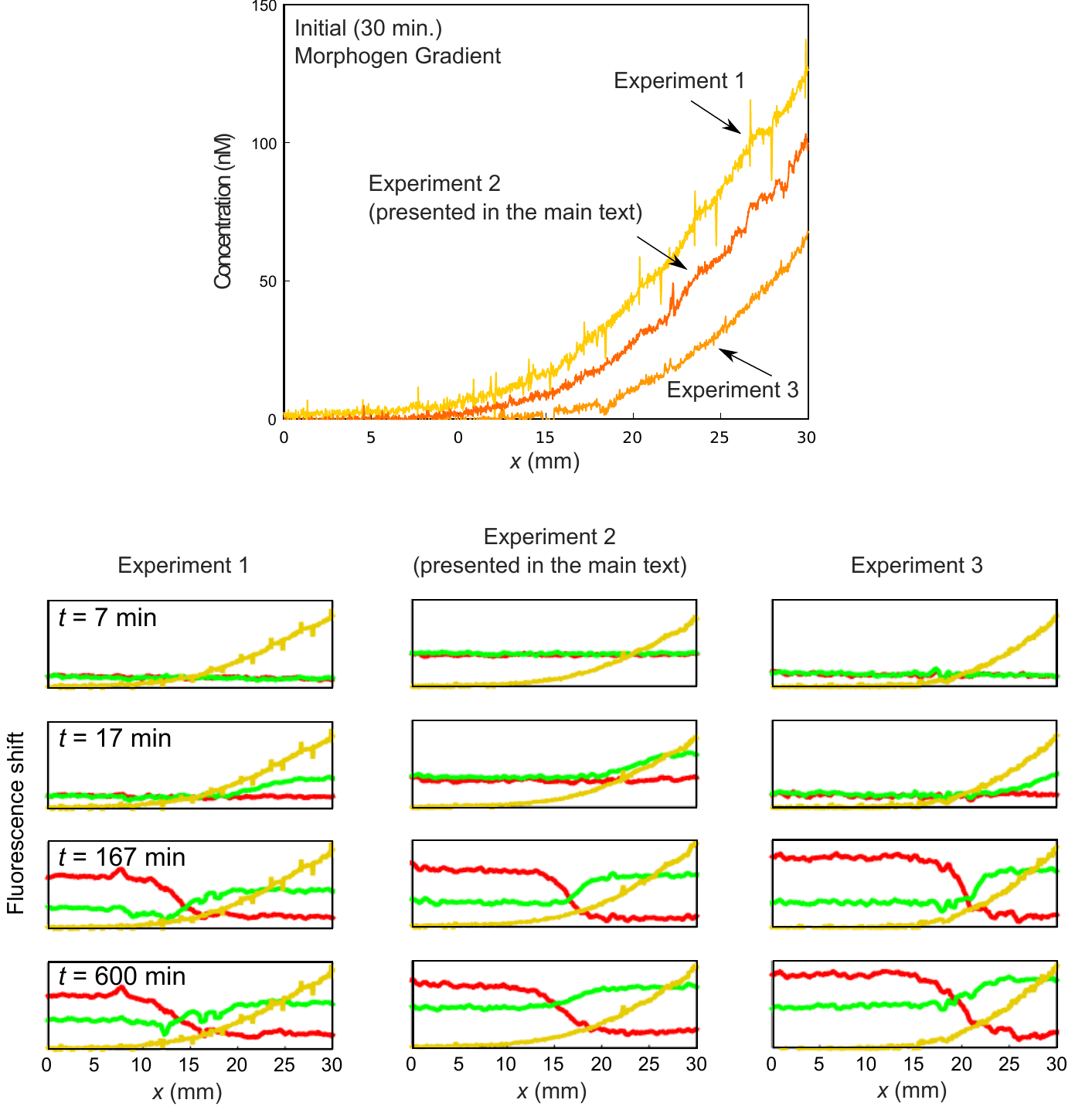}
     	\renewcommand{\figurename}{Supplementary Figure}\caption{Reproducibility of the pattern generated by the 2-species bistable network. Three experiments were performed with identical conditions, except for the three gradients of morphogen $T_{H}$ that were slightly different (top). Fluorescence shift profiles along the channel corresponding to species $H$ (green), $K$ (red) and $T_{H}$ (yellow) at different times $t$. The bumps on the yellow curves are not due to noise but to an artifact signal due to illumination inhomogeneities in each stitched field of view.}
     	\label{fig:4nodesReproducibilityA}
     \end{figure}

\begin{figure}[H]
     	\centering  
     	\includegraphics[scale=1]{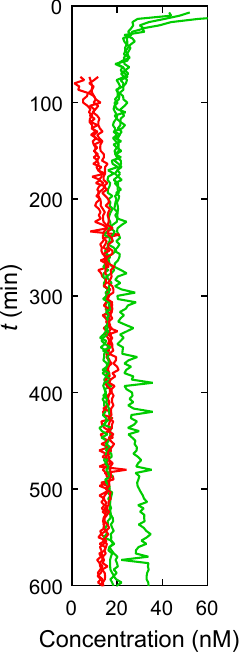}
     	\renewcommand{\figurename}{Supplementary Figure}\caption{Position of the fronts of  $H$ (green) and $K$ (red) in morphogen concentration units as a function of time for the three independent experiments in Supplementary Figure~\ref{fig:4nodesReproducibilityA}. The reproducibility is remarkable.}
     	\label{fig:4nodesReproducibilityB}
     \end{figure}

The results of three experiments with the 2-species bistable (network III) are presented to illustrate: i) their high reproducibility and ii) that shifting the morphogen gradient along $x$ shifts the position of the front by the same distance. We generated morphogen gradients of similar curvature but shifted along the $x$ axis (Supplementary Figure~\ref{fig:4nodesReproducibilityA}) for identical compositions of the reaction network. The fluorescent profiles due to species H and K were identical except for a lateral shift. The dynamics of the position of all three fronts in the morphogen concentration domain are very similar in all three experiments (Supplementary Figure \ref{fig:4nodesReproducibilityB}).

 \subsection{Immobile morphogen gradients attached to the surface \label{sec:surface}}
  
 We tested the possibility of immobilizing the morphogen gradient by attaching it to the surface using biotinylated DNA strands. Functionalization of the glass substrate was done as follows:

 \paragraph{Chemicals.} 50 mL of concentrated H$_{2}$SO$_{4}$,
 50 mL of 30\% H$_{2}$O$_{2}$, 100~$\mu$L of 100\% CH$_{3}$COOH, 150 mg of biotin-PEG-silane (Laysan Bio Inc, USA), toluene, propan-2-ol. 
 
\paragraph{Protocol.}
\begin{enumerate}
\item Prepare the 'piranha' solution by slowly adding 50 ml of 30\% H$_{2}$O$_{2}$ to concentrated H$_{2}$SO$_{4}$ with constant mixing. Note 'piranha', being both strongly acidic and oxidizing, violently reacts with organics, and is thus dangerous. 
 \item Wipe the surface of $2.5\times7.5$~mm$^2$ glass slides with propan-2-ol or acetone to get rid of majority of organics, put the glass slides into a slides box and pour 100 ml of freshly prepared 'piranha' solution. Keep for 1h.
 \item Rinse the slides by sonicating them in water several times.
 \item Prepare 100~mL of 0.44 mM solution of biotin-PEG-silane in toluene, add 100~$\mu$L of CH$_{3}$COOH and mix. Pour this mix on the glass slides to cover them and keep for 2 hours.
\item Rinse with toluene, then with propan-2-ol and finally with water and dry with nitrogen.
  \item The device finally consists of the treated glass-PEG-biotin slide, a double layer of parafilm cutted using a Graphtec CE6000 plotter cutter to define the channel geometry and a polystyrene cover slide with embedded access holes (\emph{27}). It was thermally assembled at 75°C for 5 minutes. Typical channels are $30\times2\times0.25$~mm$^3$, which corresponds to a volume of 15~$\mu$L.
  \item Before preparing the morphogen gradient, the channels were treated with streptavidin by filling them with a 1~$\mu$M solution of streptavidin and incubated at room temperature for 10 min before washing.
 Morphogen gradient profiles are then prepared by using 500 nM solution of T$_{H}$. 10 min incubation at room temperature is then necessary for DNA binding. To perform the experiment, the channels are rinsed and filled with the network solution.
\end{enumerate}
 
\begin{figure}[H]
 	\centering  
 	\includegraphics[width=0.8\linewidth]{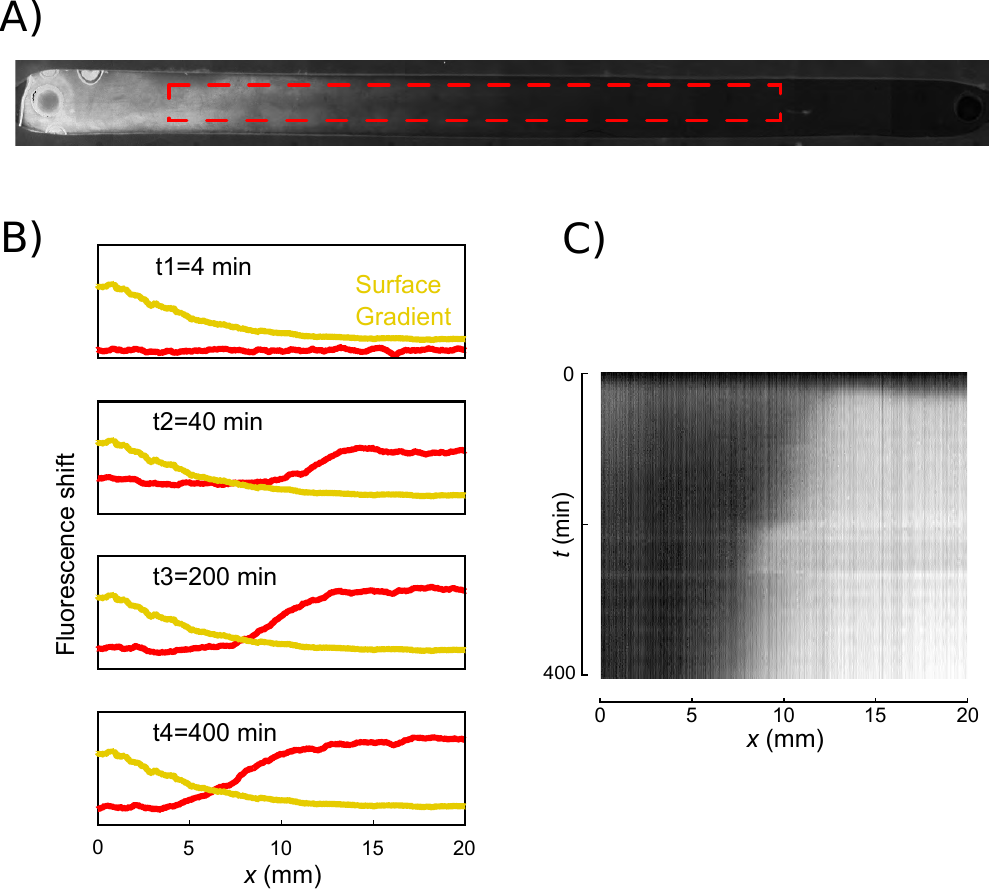}
 	\renewcommand{\figurename}{Supplementary Figure}\caption{Immobilizing the morphogen gradient onto the surface also generates an immobile concentration front. A) Fluorescence image of T$_{H}$ strand attached to a glass slide by a biotin-PEG-silane in the presence of steptavidin. The channel is carved in parafilm. B) Fluorescence profiles along the red rectangle in panel A for the morphogen (yellow) and species K (red) at different times for network III. C) Kymograph of the fluorescence associated to species K. Species concentrations for this experiment at initial time: $H=K=1$~nM, $T_{K}=20$~nM, $T_{H2R_K}=20$~nM, $T_{K2R_H}=20$~nM, the gradient on the surface was prepared using a solution of T$_{H}$  at 500~nM. }
 	\label{fig:Surface_1channel}
 \end{figure}

\begin{figure}[H]
     	\centering  
     	\includegraphics[width=\linewidth]{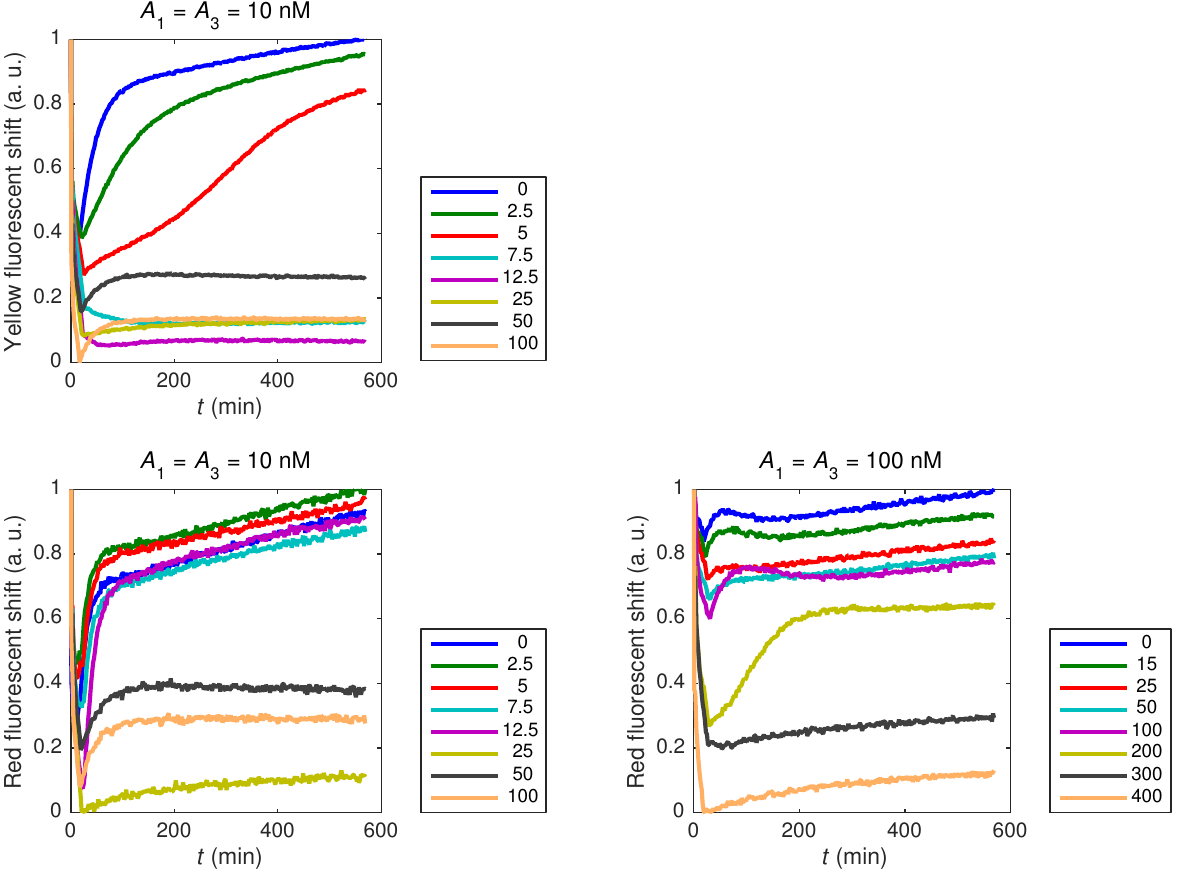}
     	\renewcommand{\figurename}{Supplementary Figure}\caption{Fluorescence \emph{vs} time for network IVb for different intial conditions ($A_1=A_3=10$~nM, left, and $A_1=A_3=100$~nM, right) and different repressor R$_2$-R$_3$ concentrations (colored lines, legend in nM). The yellow fluorescence comes from DY530 on T$_{A_3}$ and the red fluorescence from Cy3.5 on T$_{A_1}$. The yellow fluorescence shift at $A_1=A_3=100$~nM is not displayed as there was a strong cross-talk from the green channel. Both nodes of the network, A$_1$ and A$_3$ are bistable and R$_2$-R$_3$ is a bifurcation parameter. Note that at $A_1=A_3=10$~nM the A$_3$ node bifurcates at $R_2$-$R_3>5$~nM while the A$_1$ node bifurcates at $R_2$-$R_3>12.5$~nM.}
     		   	\label{fig:bifFflag}
     \end{figure}

\begin{figure}[H]
     	\centering  
     	\includegraphics[scale=0.75]{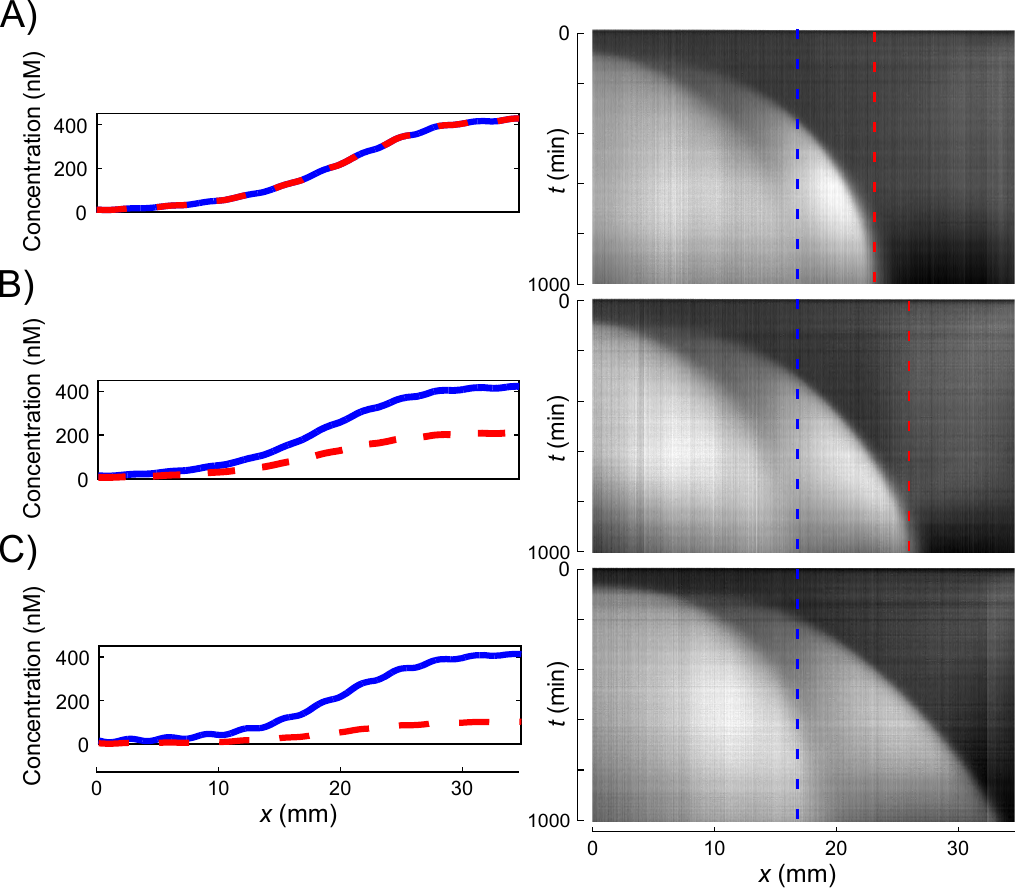}
     	\renewcommand{\figurename}{Supplementary Figure}\caption{Two orthogonal 1-species bistable networks produce a two fronts pattern which position can be independently controlled. Networks Ib and VI (Figure~S2) were introduced in a capillary. Two morphogens were used, R$_2$ and R$_3$ and gradients of similar shape but different maximal concentrations were generated. The concentration of dsDNA was followed by measuring EvaGreen fluorescence. As the two 1-species bistable networks are orthogonal, changing the morphogen gradient of one of the two networks only affects the position of the corresponding front, that moves to the right. Initial morphogen gradients (left) and EvaGreen fluorescence kymographs showing simultaneously the emergence of the two immobile fronts (right) for three different maximum values of repressor R$_2$ (red dashed lines): 400 (A), 200 (B) and 100 nM (C). In (C) the level of R$_2$ is too low to stop the front. $T_{A_2}=T_{A_3}=25$~nM, $A_2= A_3=0.5$~nM, $R_3=0-400$~nM (blue lines). $T = 45$°C.\label{fig:FFlag2pT}}
     \end{figure}

\begin{figure}[H]
 	\centering  
 	\includegraphics[width=12cm]{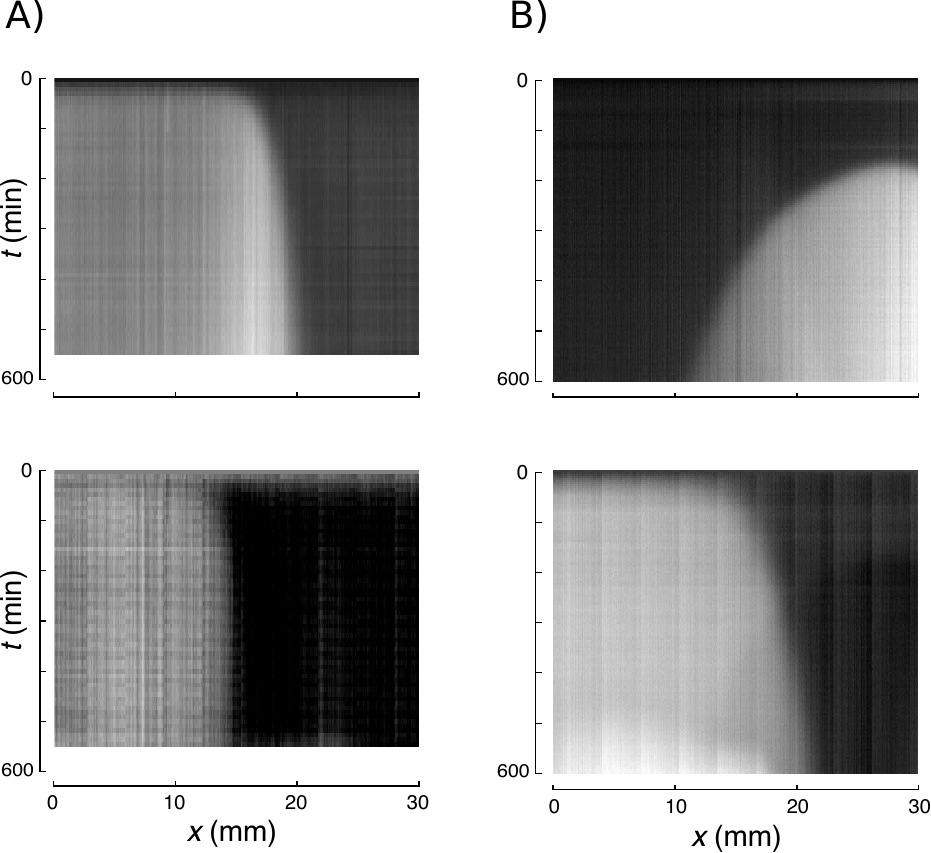}
 	\renewcommand{\figurename}{Supplementary Figure}\caption{Two French flag patterns are presented in Figure 3 of the Main Text. They come from the combination of two orthogonal 1-species bistable networks, one containing species A$_2$ and the other species A$_3$. To follow independently the concentration of A$_2$ and A$_3$ the $\textrm{T}_{A_3}$ was labeled with a yellow DY530 dye in 5'. $\textrm{T}_{A_2}$ was not labeled. It appeared that in presence of EvaGreen, only the A$_2$ system fluoresced in green, while only the A$_3$ fluoresced in yellow.  
Here we show separately the two fluorescence channels that were recorded independently during each experiment and that correspond to each bistable for the French flag patterns in Figure 3 of the Main Text. Kymographs for each color channels for the three French Flag patterns. Top row: green fluorescence, corresponding to species A$_2$. Bottom row: yellow fluorescence, corresponding to species A$_3$.\label{fig:FFlag2colors}}
 \end{figure}

\begin{figure}[H]
 	\centering  
 	\includegraphics[width=0.6\linewidth]{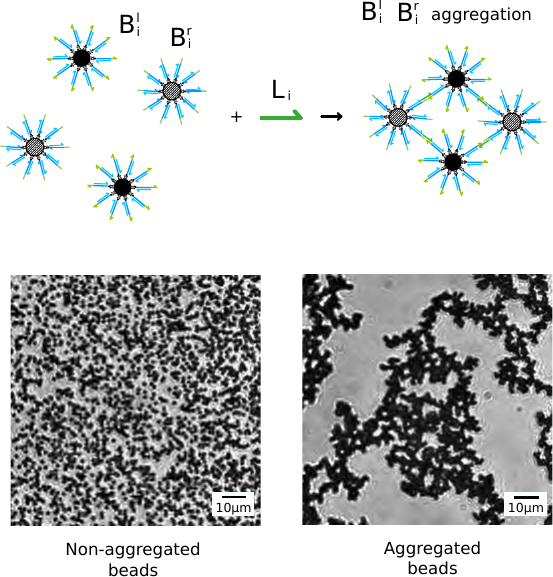}
 	\renewcommand{\figurename}{Supplementary Figure}\caption{Brightfield microscopy images show non-aggregated and aggregated beads B$_1$ after 1000 min incubation respectively in the absence (left) and in the presence of 1 $\mu$M L$_1$ (right) in the 1-species bistable buffer without enzymes at 45°C.\label{fig:BeadAggregationNoNetwork}}
 \end{figure}

\begin{figure}[H]
   	\centering  
   	\includegraphics[scale=1]{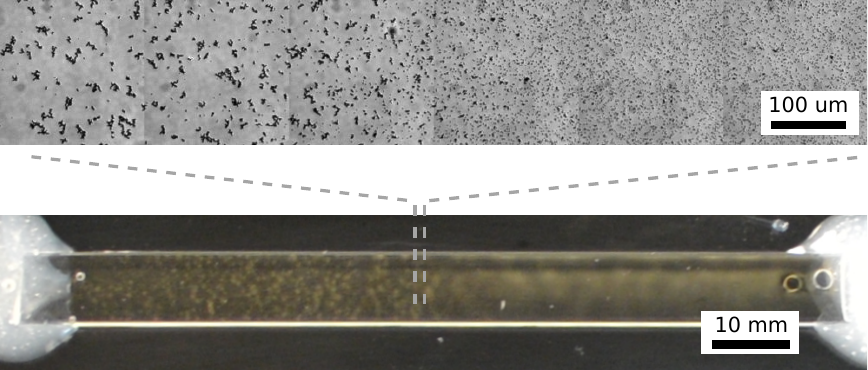}
   	\renewcommand{\figurename}{Supplementary Figure}\caption{Materialization of a Polish flag pattern with conditional B$_2$ particle aggregation. After 17~h at 45°C in the presence of network VIII, that produces linker L$_2$, we observe aggregated beads on the left and non-aggregated beads on the right. We use brightfield imaging either in high magnification with a 40$\times$ objective (top) or in low magnification, showing the entire capillary with a 35 mm lens (bottom).
 $R_3=0-400$~nM, $A_3=10$~nM, $T_{A_3}=25$~nM, $T_{L}=50$~nM, $B_2=0.2$~mg/mL}
   	\label{M161102}
   \end{figure}

\begin{figure}[H]
   	\centering  
   	\includegraphics[scale=1]{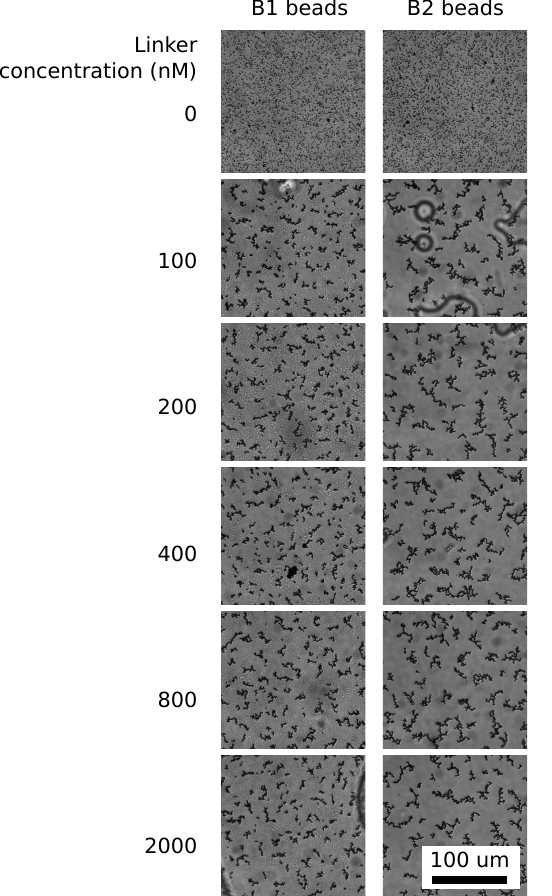}
   	\renewcommand{\figurename}{Supplementary Figure}\caption{Titration of  two separate solutions, each containing beads B$_1$ or B$_2$ with their respective linkers L$_1$ and L$_2$, in the absence of the networks. In the absence of linker, aggregation is not observed. 100~nM of linker is enough for a full aggregation. Each brightfield image covers a field of view of 0.2~mm and is taken using a 40x objective. B$_1$ beads images have been recorded after 2~h at 45°C, B$_2$ beads images have been recorded after 20~min at 45°C. This difference in time is not significant but related to experimental constraints. Conditions: beads were at 0.5 mg/mL in the 1-species bistable buffer.}
   	\label{M161010_161102}
   \end{figure}

\begin{figure}[H]
     	\centering  
     	\includegraphics[width=\linewidth]{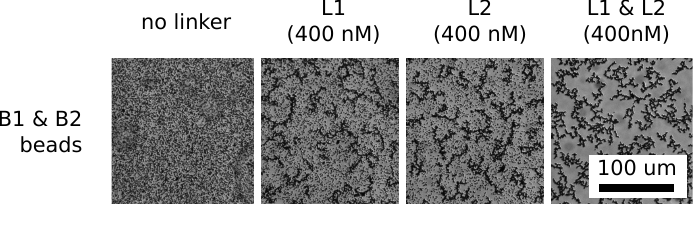}
     	\renewcommand{\figurename}{Supplementary Figure}\caption{Orthogonality of the linkers in the absence of the networks. A solution containing the two types of beads B$_1$ and B$_2$ was maintained at 45°C for 2~h in the presence of, from left to right: no linker, 400~nM L$_1$ linker, 400~nM L$_2$ linker, 400~nM L$_1$ and L$_2$ linkers.  In the absence of linker, aggregation is not observed while when both linkers are present, both type of beads aggregate. When only L$_1$ or L$_2$ is present, half of beads aggregate, the others remaining Brownian. Each brightfield image covers a field of view of 0.2~mm and is taken using a 40$\times$ objective.
     			$B_1=B_2= 0.5$~mg/mL. 
     		}
     	\label{M161019}
     \end{figure}

\begin{figure}[H]
      	\centering  
      	\includegraphics[scale=1]{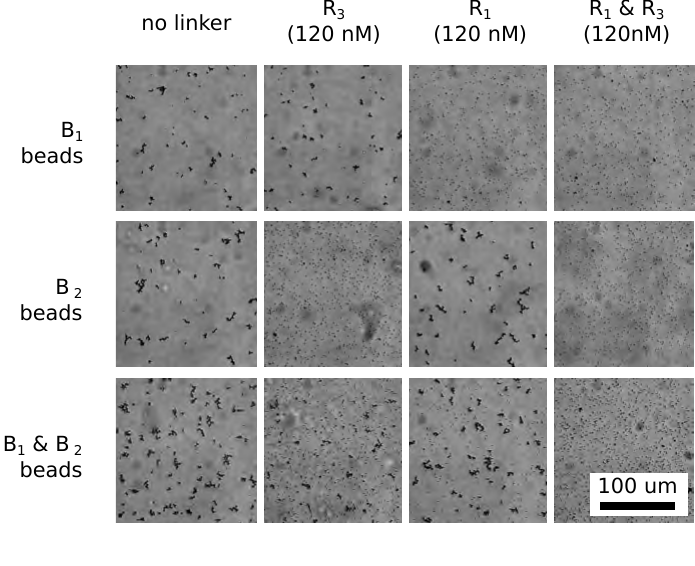}
      	\renewcommand{\figurename}{Supplementary Figure}\caption{Orthogonality of bead pairs B$_1$ and B$_2$ in the presence of the French flag networks VII and VIII. Three solutions containing B$_1$ beads, B$_2$ beads, and B$_1$ together with B$_2$ beads were prepared containing both networks VII and VIII. For each solution, four conditions were studied: in the absence of both repressors, in the presence of 120~nM R$_1$ repressor, in the presence of 120~nM R$_3$ repressor and in the presence of both R$_1$ and R$_3$ repressors.
      		In the absence of repressor (left column), aggregation is observed in all three bead solutions, while when both repressors are present (right column), beads do not aggregate in any of these solutions. In the solution containing B$_1$ beads (top row), aggregation is observed only when R$_1$ is absent. In return, in the solution containing B$_2$ beads (center row), aggregation is observed only when R$_3$ is absent. When both B$_1$ and B$_2$ beads are present (bottom row), the presence of one repressor, either R$_1$ or R$_3$, leads to partial aggregation. In summary, the beads respond to the networks as designed. Each brightfield image covers a field of view of 0.2~mm and is taken using a 40$\times$ objective.
  		$R_1=R_3=120$~nM, $A_1=A_3=10$~nM, $T_{A_1}=T_{A_3}=25$~nM, $T_{L1}=T_{L2}=50$~nM. Images recorded at 45°C, either after 19h (top and middle rows) or after 3~h (bottom row). All beads were used at a concentration of 0.1~mg/mL.	}
      	\label{M161107_161109}
      \end{figure}

\begin{figure}[H]
        	\centering  
        	\includegraphics[scale=0.45]{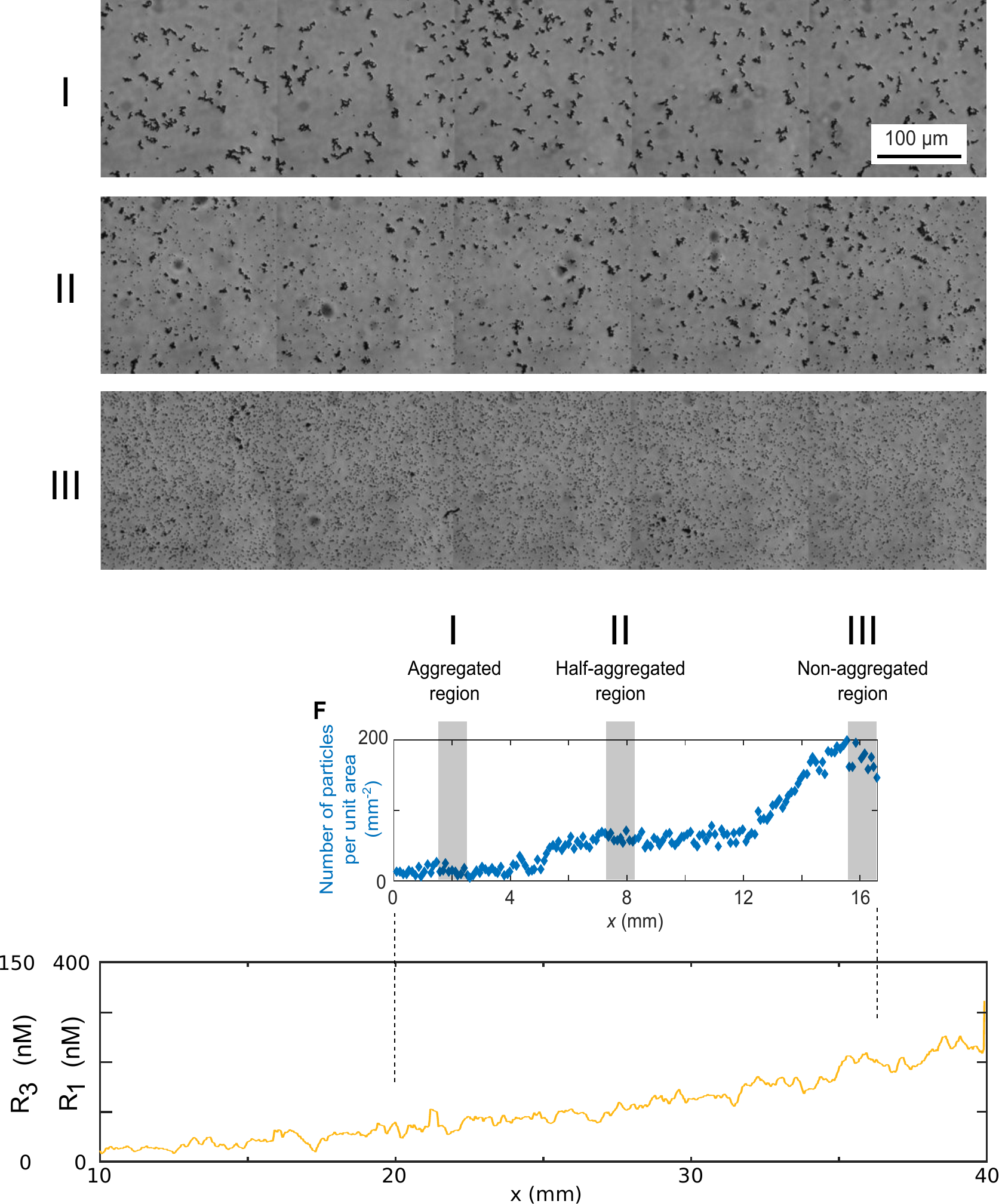}
        	\renewcommand{\figurename}{Supplementary Figure}\caption{Materialization of a French flag pattern with conditional B$_1$ and B$_2$ bead aggregation. After 19~h at 45°C, in the presence of networks VII and VIII that produce linker L$_1$ and L$_2$ respectively. The bead aggregation is clearly seen on the left hand side of the capillary while no aggregation occurs on the right hand side. In the center of the capillary, corresponding  to a region of space in which only the linker L$_2$ is produced, only one type of beads aggregates (B$_2$), the second type of beads remaining Brownian. Conditions:
        			$R_1=0-400$~nM, $R_3=0-150$~nM, $A_1=A_3=10$~nM, $T_{A_1}=T_{A_3}=25$~nM, $T_{L1}=T_{L2}=50$~nM, 
        			$B_1=B_2=0.1$~mg/mL, images recorded after 19~h at 45°C. Same experiment as in Figure~\ref{fig4} in the MT.}
        	\label{M161108}
        \end{figure}

Immobile morphogen gradients generated Polish flag patterns with network III (Supplementary Figure~\ref{fig:Surface_1channel}). However, it was difficult to obtain reproducible gradients on the surface. For this reason and because gradients in solution were very robust they were preferred throughout this work. 
 
  \section{Data associated to Figure 3 in the Main Text}
 \subsection{Bifurcation of network IVb in a well-mixed reactor}

To be able to observe the extent of reaction of each node independently in temporal kinetic experiments we modified slightly the network. In Figure 3A of the Main Text we used network IVa (with A$_2$ growing on non-fluorescently labeled T$_{A_2}$) while here we used network IVb (with A$_1$ growing on T$_{A_1}$ fluorescently labeled with Cy3.5). A$_1$ and A$_2$ only differ by one base. 
  
 \subsection{The position of the borders of the French flag pattern can be independently controlled}
        
 \section{Data associated to Figure 4 in the Main Text}

   \subsection{Polish flag with B$_2$ beads}

   \subsection{Titration of the aggregation of the two pairs of beads B$_1$ and B$_2$ with their linkers L$_1$ and L$_2$}

     \subsection{Orthogonality of the linkers}

      \subsection{French flag network with beads in well-mixed conditions}

        \subsection{Materialization of a French flag pattern with conditional B$_1$ and B$_2$ particle aggregation}

      \begin{figure}[H]
      	\centering  
      	\includegraphics[scale=1]{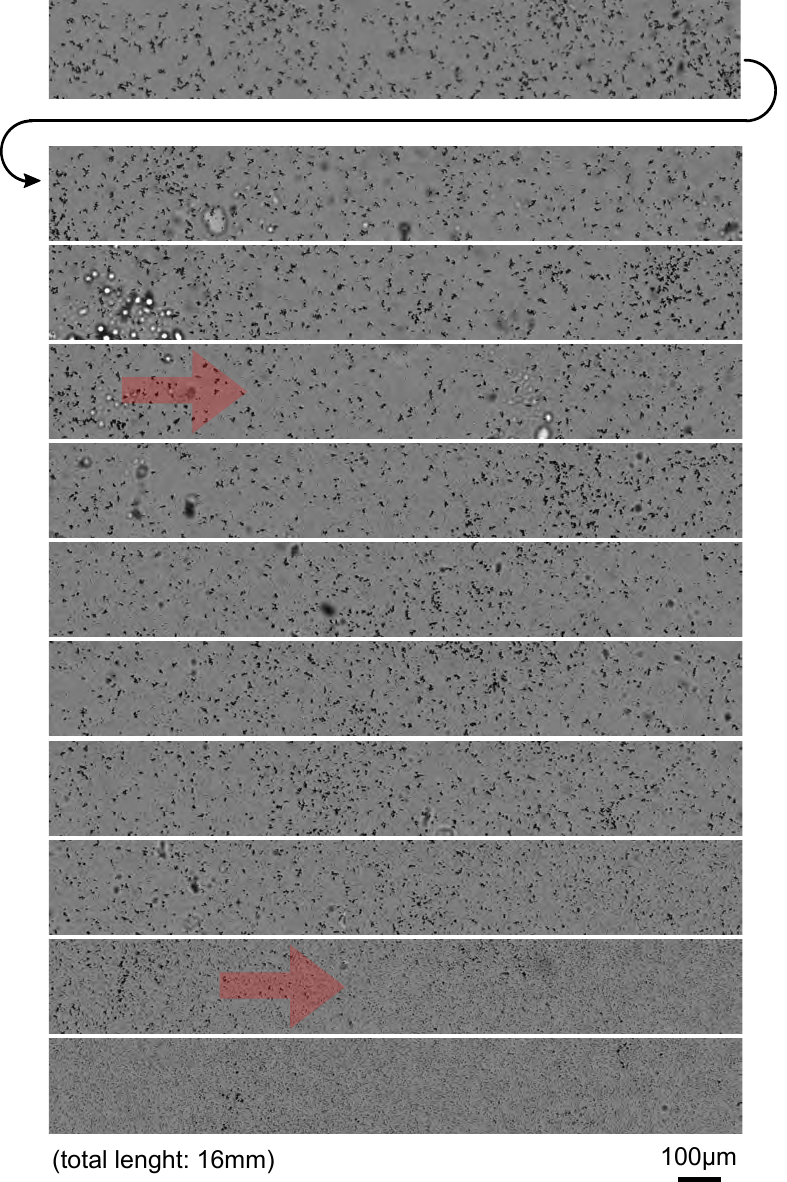}
      	\renewcommand{\figurename}{Supplementary Figure}\caption{Whole-capillary image of the aggregation French flag pattern. 77 contiguous fields of view representing 16~mm along the capillary taken using a 40$\times$ objective.
      		On top (left hand side of the capillary)  aggregation of all beads is clearly observed while at the bottom (right hand side) no aggregation occurs. In between, over 7~mm, only one type of beads aggregated. Transition zones are indicated by red arrows. Background has been subtracted on this image. Same experiment as in Figure~\ref{fig4} in the MT.}
      	\label{M161108fullgradient}
      \end{figure} 
  
 \subsection{Long-term stability of the aggregated bead pattern}
  
          \begin{figure}[H]
         	\centering  
         	\includegraphics[width=10cm]{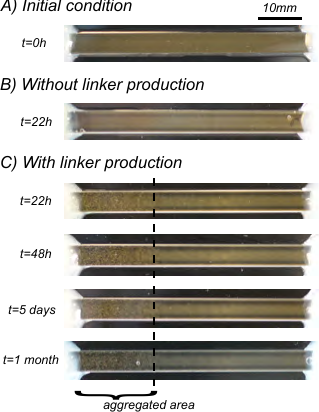}
         	\renewcommand{\figurename}{Supplementary Figure}\caption{The aggregation of beads under the control of a Polish flag pattern is irreversible and visible to the naked eye. A) At $t=0$, the dispersion of beads B$_1$ in the capillary appears as a brown homogeneous background. B) After 22~h, in the presence of network I, that does not produce linker L$_1$, the beads sediment homogeneously along the longitudinal axis of the capillary but away from the walls and do not aggregate. C) In contrast, in the presence of network VII, that produces linker L$_1$, the bead aggregation is clearly seen on the left hand side of the capillary. The pattern of bead aggregation is stable for at least one month. The photos were taken in brightfield with a Nikon D600 camera equipped with a 35~mm lens.\label{fig:morphoflag_byEyes}}
         \end{figure}

\newpage

\section{Data associated to the Discussion in the Main Text}

\subsection{Dependence of the front width on $D$ and on the thresholded degradation kinetics}

Dimensional analysis indicates that a reaction-diffusion process governed by a diffusion coefficient $D$ and a first-order kinetic rate $k$ should generate a pattern with a characteristic length $\lambda\sim\sqrt{D/k}$. We thus expect that the width of the Polish flag pattern is given by $\lambda$.

To illustrate this intuition we used a simple one-variable model of the 1-species bistable\cite{Montagne2016}

\begin{equation}
\frac{\partial A(x,t)}{\partial t}  = A(x,t) \left( \frac{k_p}{K_p+A(x,t)} -\frac{k_d R(x)}{K_dR(x) + A(x,t)} - \frac{k_r}{K_r+A(x,t)}\right) + D \frac{\partial^2 C(x,t)}{\partial x^2} + \epsilon \label{simuModelPt}
\end{equation}
where $A(x,t)$ is the concentration of the autocatalytic species that is produced with rate $k_p$ and saturation $K_p$, repressed by the gradient of repressor $R(x)$ at rate $k_d$ with saturation $K_d$ and degraded directly by the exonuclease with rate $k_r$ and saturation $K_r$. We integrated this adimensional equation with $k_p =1 $, $k_r = 30$, $K_p =1$, $K_d= 0.1$ and $K_r =100$ for $x\in[0,10]$ and for different values of $k_d$ and $D$ in the presence of a gradient $R(x) = 10e^{(x - 10)}$. We obtained Polish flag-type patterns at steady state which profiles are shown in Supplementary Figure~\ref{fig:pTSimu}AB. We extracted the width of these profiles by fitting a decaying exponential $e^{-(x-x_0)/\lambda}$. As expected from the scaling law, $\lambda^2$ depended linearly on both $D$ and $1/k_d$ (Supplementary Figure~\ref{fig:pTSimu}C-D).

\begin{figure}[H]
 	\centering  
 	\includegraphics[width=\linewidth]{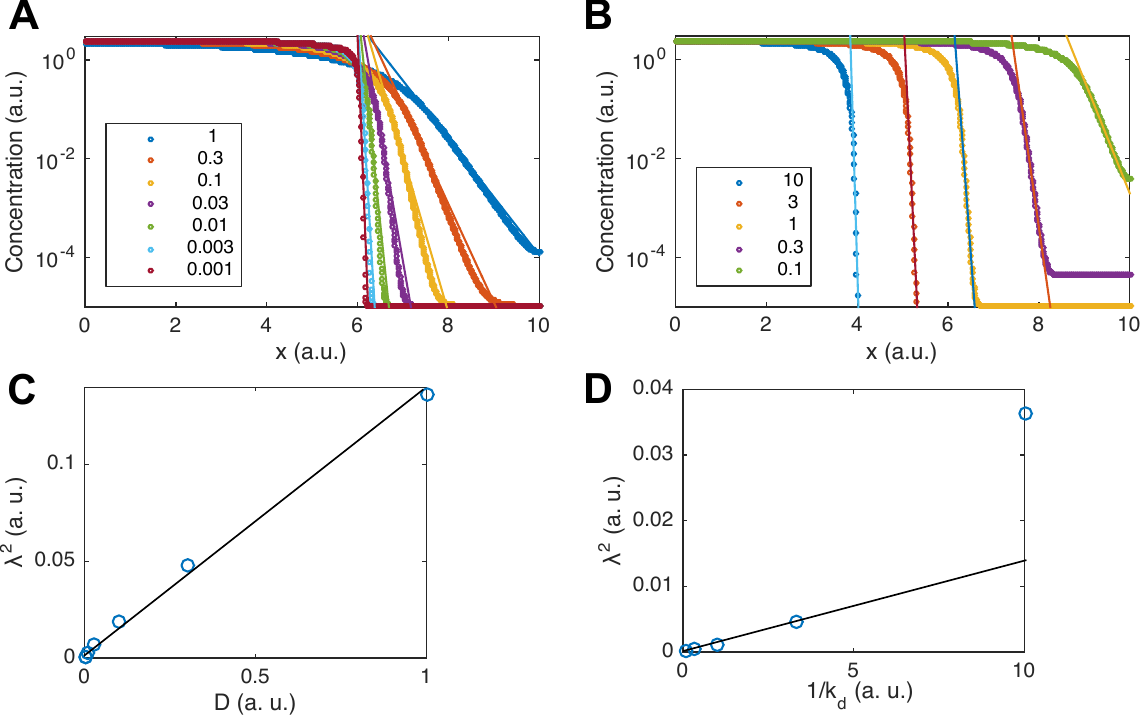}
 	\renewcommand{\figurename}{Supplementary Figure}\caption{Dependence of the steady-state solutions of Equation~\ref{simuModelPt} on $D$ and $k_d$. Steady-state concentration profiles for different values of $D$ (A) and $k_d$ (B). Dots correspond to data from the simulations, lines correspond to exponential fits. Squared front width, $\lambda$, vs. $D$ (C) and $k_d$ (D), black lines are linear fits.\label{fig:pTSimu}}
 \end{figure}

\subsection{Degradation kinetics of species W$_2$ in the presence of R$_2$ and exonuclease}

\begin{figure}[H]
 	\centering  
 	\includegraphics[width=10cm]{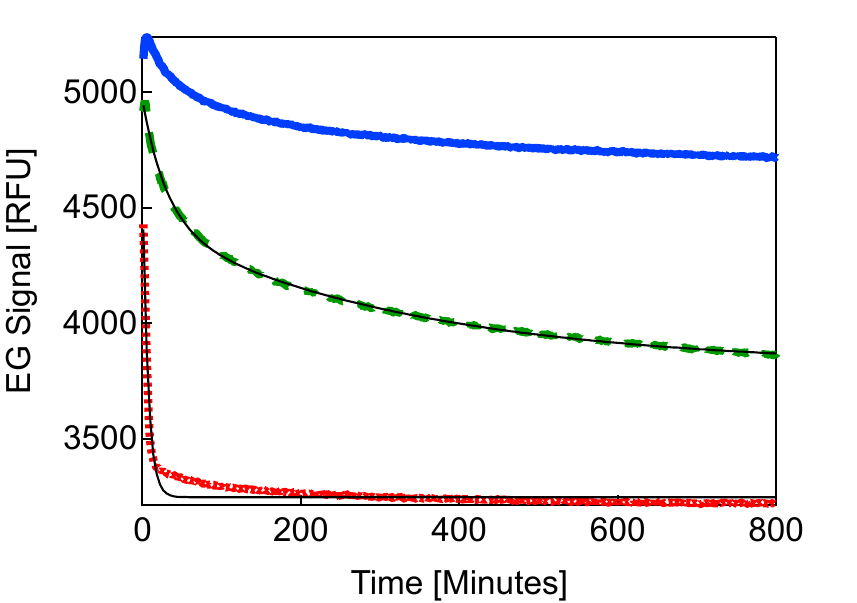}
 	\renewcommand{\figurename}{Supplementary Figure}\caption{Degradation of W$_2$ in the presence of R$_2$ and exonuclease. EvaGreen fluorescence vs. time for solutions of R$_2$ + W$_2$ in the absence (blue line) and in the presence (green long dashed line) of exonuclease, and for a solution of R$_2$ + $A_2$ in the presence of exo (red short dashed line). Thin black lines are fits to the data, on the green curve a biexponential fit ($3813.4 + 618.83\exp(-0.0030224t) + 530\exp(-0.032745t)$) and on the red curve a monoexponential fit ($3245.9+ 1342.7\exp(-0.14582t)$). Note that the degradation rate of $A_2$ is almost 100 times faster than the slowest rate of degradation of $W_2$, suggesting that an optimization of this reaction could result in significantly faster rates that would increase the resolution of the patterns. Each oligonucleotide was present at 200 nM. ttRecJ exonuclease was used at 31.25 nM (2.5\%), 2.5-fold more concentrated than in patterning experiments. \label{fig:W1degradation}}
 \end{figure}

\newpage
 
%
%
%
%




   \setcounter{figure}{0} 
    
\newpage
\section{Supplementary movies}

\subsection{Movie S1 -  A DNA-based network with two self-activating nodes that repress each other generates two immobile fronts that repel each other}
      
  \begin{figure}[H]
  	\centering  
  	\includegraphics[scale=0.4]{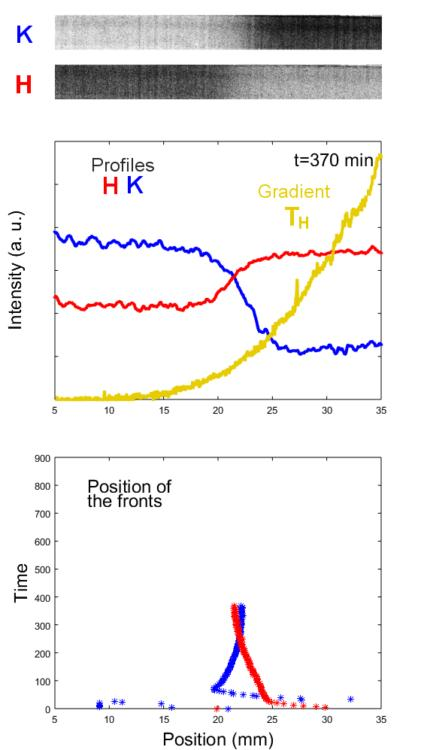}
  	\renewcommand{\figurename}{Movie S}
	\caption{
Dynamics of the 2-species bistable (network III) in a gradient of  T$_{H}$. Top: Fluorescence images of the green and red channels associated to the concentration of species H and K within the channel. Middle: Fluorescence profiles along the length of the channel (averaged across its width) for H (red), K (blue) and morphogen T$_{H}$ (yellow) at different times. Bottom: Position of the fronts of H and K extracted from  the middle panel. Movie associated to Figure 2 in the Main Text.
}
 	\label{fig:Movie1}
  \end{figure}

           \newpage

   \subsection{Movie S2 - Synthesis of a reaction-diffusion French Flag pattern}

      \begin{figure}[H]
      	\centering  
      	\includegraphics[scale=0.4]{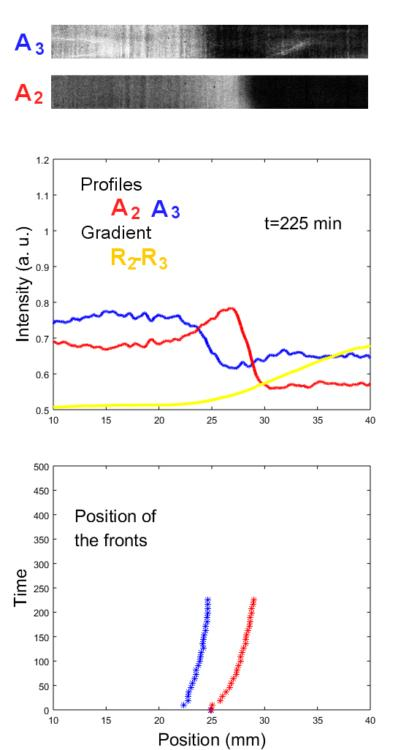}
      	\renewcommand{\figurename}{Movie S}\caption{Dynamics of the coupled 1-species bistables (network IVa) in a gradient of R$_2$-R$_3$ $0-400$~nM. Top: Fluorescence images of the green and yellow channels associated to the concentration of species A$_2$ and A$_3$ within the channel. Middle: Fluorescence profiles along the length of the channel (averaged across its width) for A$_2$ (red), A$_3$ (blue) at different times, and initial morphogen R$_2$-R$_3$ profile (yellow). Bottom: Position of the fronts of A$_2$ and A$_3$ extracted from the middle panel. Movie associated to Figure 3A in the Main Text.}
      	\label{fig:Movie2}
      \end{figure} 
    

%
%
%
%

%
%
%
%
%
%
%
%
%
%



\end{document}